\documentclass[a4paper,11pt]{article}
\pdfoutput=1 % if your are submitting a pdflatex (i.e. if you have
\usepackage{jcappub} % for details on the use of the package, please
\usepackage[utf8]{inputenc}

\usepackage{siunitx}
\usepackage{framed}
\usepackage{lipsum}
\usepackage{xcolor}
\usepackage{hyperref}
\usepackage{graphicx}% Include figure files
\usepackage{float}
\usepackage{dcolumn}% Align table columns on decimal point
\usepackage{bm}% bold math
\usepackage{comment} % multiline comments. Before submssion we shoult take them out
\usepackage{subcaption}

\usepackage{booktabs}
\usepackage{bbold}

\usepackage{placeins} % for \FloatBarrier

\def\LALO          {LaAlO$_3$}

\title{\vspace{-0.6em}Simulating MADMAX in 3D: Requirements for Dielectric Axion Haloscopes \\\vspace{-0.4em}}

\subheader{\vspace{-2.5em}\\\hbox to\textwidth{\hfill MPP-2021-61}}

\newcommand{\mpp}{a}
\newcommand{\uhh}{b}
\newcommand{\cppm}{c}
\newcommand{\cnrs}{d}
\newcommand{\mpira}{e}
\newcommand{\utue}{f}
\newcommand{\desy}{g}
\newcommand{\rwth}{h}
\newcommand{\uzz}{j}

\author[\mpp,1,*]{S.~Knirck,\note{Now at Fermi National Accelerator Laboratory, Batavia, IL 60510, USA}}
\author[\uhh,2,*]{J.~Sch\"utte-Engel,\note{Now at University of Illinois at Urbana-Champaign, Urbana, IL 61801, USA}\note[*]{Corresponding authors.}}
\author[\cppm]{S.~Beurthey,}
\author[\uhh]{D.~Breitmoser,}
\author[\mpp]{A.~Caldwell,}
\author[\cppm]{C.~Diaconu,}
\author[\mpp]{{J.~Diehl},}
\author[\uhh]{J.~Egge,}
\author[\cnrs]{M.~Esposito,}
\author[\uhh]{A.~Gardikiotis,} 
\author[\uhh]{E.~Garutti,}
\author[\mpira]{S.~Heyminck,}
\author[\cppm]{F.~Hubaut,}
\author[\utue]{J.~Jochum,}
\author[\cppm]{P.~Karst,}
\author[\mpira]{M.~Kramer,}
\author[\uhh]{C.~Krieger,}
\author[\cppm]{D.~Labat,}
\author[\mpp]{C.~Lee,}
\author[\mpp]{X.~Li,}
\author[\desy]{A.~Lindner,}
\author[\mpp]{B.~Majorovits,}
\author[\uhh]{S.~Martens,} 
\author[\uhh]{M.~Matysek,} 
\author[\rwth]{E.~\"Oz,}
\author[\cnrs]{L.~Planat,}
\author[\cppm]{P.~Pralavorio,}
\author[\mpp]{G.~Raffelt,}
\author[\cnrs]{A.~Ranadive,}
\author[\uzz]{J.~Redondo,}
\author[\mpp]{O.~Reimann,}
\author[\desy]{A.~Ringwald,}
\author[\cnrs]{N.~Roch,}
\author[\desy]{J.~Schaffran,}
\author[\rwth]{A.~Schmidt,}
\author[\mpp]{L.~Shtembari,}
\author[\mpp]{F.~Steffen,}
\author[\utue]{C.~Strandhagen,} 
\author[\mpp]{D.~Strom,}
\author[\utue]{I.~Usherov,}
\author[\mpira]{G.~Wieching}
\author[]{(MADMAX collaboration)}

\affiliation[\mpp]{Max Planck Institute for Physics, 80805 Munich, Germany}
\affiliation[\uhh]{Universit\"at Hamburg, 22761 Hamburg, Germany}
\affiliation[\cppm]{Centre de Physique des Particules de Marseille, Aix Marseille Univ, CNRS/IN2P3, CPPM, Marseille, France}
\affiliation[\cnrs]{Univ. Grenoble Alpes, CNRS, Grenoble INP, Institut Neel, 38000 Grenoble, France}
\affiliation[\mpira]{Max Planck Institute for Radio Astronomy, 53121 Bonn, Germany}
\affiliation[\utue]{Eberhard Karls Universit\"at T\"ubingen, 72074 T\"ubingen, Germany}
\affiliation[\desy]{Deutsches Elektronen-Synchrotron, 22607 Hamburg, Germany}
\affiliation[\rwth]{RWTH Aachen University, 52062 Aachen, Germany}
\affiliation[\uzz]{Universidad de Zaragoza, 50009 Zaragoza, Spain}

\emailAdd{knirck@fnal.gov}
\emailAdd{jans@illinois.edu}

\abstract{We present 3D calculations for dielectric haloscopes such as the currently envisioned MADMAX experiment. For ideal systems with perfectly flat, parallel and isotropic dielectric disks of finite diameter, we find that a geometrical form factor reduces the emitted power by up to 30\,\% compared to earlier 1D calculations. We derive the emitted beam shape, which is important for antenna design. We show that realistic dark matter axion velocities of $10^{-3} c$ and inhomogeneities of the external magnetic field at the scale of $10\,\%$ have negligible impact on the sensitivity of MADMAX. We investigate design requirements for which the emitted power changes by less than 20\,\% for a benchmark boost factor with a bandwidth of $50\,{\rm MHz}$ at $22\,{\rm GHz}$, corresponding to an axion mass of $90\,\mu{\rm eV}$. We find that the maximum allowed disk tilt is $100\,\mu{\rm m}$ divided by the disk diameter, the required disk planarity is $20\,\mu{\rm m}$ (min-to-max) or better, and the maximum allowed surface roughness is $100\,\mu{\rm m}$ (min-to-max). We show how using tiled dielectric disks glued together from multiple smaller patches can affect the beam shape and antenna~coupling.}

\begin{document}

\maketitle

\flushbottom

\section{Introduction}
\label{sec:intro}

The QCD axion arises naturally as a solution of the strong CP problem of the Standard Model~(SM) 
of particle physics~\cite{PhysRevLett.38.1440,PhysRevLett.40.223,PhysRevLett.40.279}. It is furthermore one of the most appealing candidates for cold dark matter (CDM)~\cite{DINE1983137,PRESKILL1983127,ABBOTT1983133,DAVIS1986225,LYTH1992189,Kawasaki:2014sqa,Fleury:2015aca,Ringwald:2015dsf,Fleury:2016xrz,Borsanyietal,Ballesteros:2016xej}
with a viable mass range from $m_a=\SI{e-13}{\electronvolt}$ to $\SI{e-2}{\electronvolt}$~\cite{PhysRevD.98.030001}.
Depending on the axion-photon coupling $g_{a\gamma}$, cavity experiments have excluded a small mass range around a few~\si{\micro\electronvolt}~\cite{Braine:2019fqb,Backes:2020ajv}, with a huge range in mass remaining unprobed. 
Axion masses larger than $\SI{26}{\micro\electronvolt}$ are motivated by the post-inflationary Peccei-Quinn symmetry breaking scenario~\cite{Kawasaki:2014sqa,Fleury:2015aca,PhysRevD.85.105020,Klaer:2017ond,Kawasaki:2018bzv,Gorghetto:2018myk,Borsanyietal,Ringwald:2015dsf,PhysRevD.98.030001}. The MADMAX (MAgnetized Disk and Mirror Axion eXperiment) dielectric haloscope~\cite{millar2017dielectric,TheMADMAXWorkingGroup:2016hpc,Brun:2019lyf} is one of the most ambitious initiatives targeting the axion mass range around $\SI{100}{\micro\electronvolt}$.
For other experiments in this and other mass ranges,
see~\cite{Horns:2012jf,TheMADMAXWorkingGroup:2016hpc,McAllister:2017lkb,Jeong:2017hqs,Melcon:2018dba,Baryakhtar:2018doz,Lawson:2019brd,Suzuki:2015sza,FUNK:2017icw,Knirck:2018ojz,BRASS:website,Raaijmakers:2019hqj,PhysRevLett.123.121601,Zarei:2019sva,10.1007/978-3-319-92726-8_16,BARBIERI2017135,PhysRevX.4.021030,Geraci:2017bmq,Melcon:2018dba,Semertzidis:2019gkj,Schutte-Engel:2021bqm}, for a review cf.~\cite{PhysRevD.98.030001,Graham:2015ouw,Irastorza:2018dyq}.

In the presence of a strong external $B$-field, axions are converted into electromagnetic radiation at 
interfaces of media with different dielectric constants $\epsilon$. 
The MADMAX experiment consists of a metallic mirror and many parallel dielectric disks in vacuum leading to electromagnetic radiation from each interface separating regions with different $\epsilon$. Depending on the disk positions the radiation from different interfaces can interfere constructively and excite resonances between the dielectric disks, although with significantly lower quality factors as cavity experiments.
The power boost factor~$\beta^2$ describes the enhancement of the power emitted by the mirror together with the set of dielectric disks (booster) with respect to the radiation emitted by a perfect mirror of the same area and under the same $B$-field. 
Previous one dimensional (1D) calculations~\cite{millar2017dielectric} showed that with 80 lanthanum aluminate (\LALO{}) disks ($\epsilon \approx 24$) a power boost factor of $\approx \num{5e4}$ can be achieved over a bandwidth of \SI{50}{\mega\hertz}, leading to an emitted power of
    \begin{equation}
    	{P_\gamma}
    	= 1.6\times 10^{-22}\,\si{\watt} \left(\frac{\beta^2}{5\times 10^4}\right) 
    	\left(\frac{A}{\SI{1}{\square\metre}}\right) \left(\frac{B_{\rm e}}{10~{\rm T}}\right)^2 \, \left(\frac{|C_{a\gamma}|}{1}\right)^2 \,  \left(\frac{\rho_a}{\SI{0.45}{\giga\electronvolt\per\centi\meter\cubed}}\right) ,
	\end{equation}
where $A$ is the surface of the dielectric disks, $B_{\rm e}$ the strength of the external magnetic field, $\rho_a$ the local cold dark matter density and $|C_{a\gamma}|$ a model-dependent coupling constant proportional to the axion-photon coupling $g_{a\gamma}$ as defined in~\cite{millar2017dielectric}, with typical values of $|C_{a\gamma}| \approx 1.9$ (KSVZ model~\cite{Kim:1979if,Shifman:1979if}) or $|C_{a\gamma}| \approx 0.7$ (DFSZ model~\cite{Dine:1981rt,Zhitnitsky:1980tq}).

It is of central importance to understand the systematic uncertainties in the power boost factor $\beta^2$. Previous work has relied on a 1D model for $\beta^2$~\cite{millar2017dielectric}, while three dimensional (3D) effects have only been taken into account for smaller systems with up to one dielectric disk~\cite{Knirck:2019eug,Schutte-Engel:2018mfn}. The work presented here extends these studies to systems with multiple dielectric disks as envisioned for MADMAX. We present simulations taking some of the most important realistic boundary conditions for an open booster (disks surrounded by free space) into account, i.e., first of all the fact that the disks are of finite size (ideal 3D booster), but also implications from a finite axion velocity, magnetic field inhomogeneities,  mechanical tolerances, imprecise disk geometries,  tilts and tiled disks (non-ideal booster).
To this end we apply the finite element method by using the azimuthal symmetry of the booster (2D3D~FEM), as well as the Recursive Fourier Propagation method, both introduced in~\cite{Knirck:2019eug}. In addition, we use the Mode Matching formalism briefly described in the next chapter. For a comparison showing their consistency see appendix~\ref{app:comparision}.

The paper in large parts is based on results from two PhD theses~\cite{Knirck:2020,SchtteEngel:450146}. It is structured as follows: In section~\ref{sec:modes} we identify eigenmodes independently propagating inside the system, which form the basis for our description of the booster.
Section~\ref{sec:ideal_system} deals with finite-diameter but perfectly parallel and flat dielectric disks  to which we refer as the ideal 3D case. Finally in section~\ref{sec:nonideal_system} we study non-ideal effects including effects from a~finite axion velocity, $B$-field inhomogeneities, disk tilts and surface inaccuracies. We also discuss dielectric disks glued together from smaller uniform patches (tiled disks).

\section{System Modes}\label{sec:modes}

At first order in the axion-photon coupling $g_{a\gamma}$ the axion-Maxwell equations can be written (using natural units with $\hbar = c = 1$ and the Lorentz-Heaviside convention
$\alpha=e^2/4\pi$) as a wave equation for the electric field $\bm{E}$ using time-harmonic %
fields as~\cite{Knirck:2019eug}
\begin{eqnarray}
-\nabla^2 \bm{E} + \nabla ( \nabla \cdot \bm{E} ) - \epsilon \omega^2 \bm{E}
= \omega^2 \, g_{a\gamma} \bm{B}_{\rm e} a,
\label{eq:sim:3d:waveequation-td}
\end{eqnarray}
where $\omega = 2\pi\nu = m_a$ is the angular frequency and $\epsilon$ is the permittivity. The permeability is assumed to be $\mu = 1$.
The axion field $a$ on the right hand side acts as a source of electric fields, through its coupling constant $g_{a\gamma}$ and external magnetic field $\bm{B}_{\rm e}$. For $\epsilon$ and $\bm{B}_{\rm e}$ constant over lengths much larger than the free photon wavelength~$\lambda = 2\pi/\omega$~\cite{Redondo:2010js,Ouellet:2018nfr,Knirck:2019eug}, a solution is given by the axion-induced field
\begin{equation}
    \bm{E}_a = - \frac{g_{a\gamma} \bm{B}_{\rm e} a}{\epsilon}.
\end{equation}
The axion-induced field $\bm{E}_a$ has a discontinuity at a boundary between regions with different $\epsilon$ and hence does not solve eq.~\eqref{eq:sim:3d:waveequation-td} anymore. The full solution is obtained by adding emitted electromagnetic radiation from the boundary compensating the discontinuity~\cite{millar2017dielectric}.
In all figures throughout this paper the electric fields are shown in units of $E_0 \equiv \max |\bm{E}_a|$ at a fixed instant of time if not stated otherwise.

For simplicity (and when not using FEM methods as e.g.\ in section~\ref{sec:tiling}), we will neglect free charges in the following by setting $\nabla \cdot \bm{E} = 0$. This sets the second term in eq.~\eqref{eq:sim:3d:waveequation-td} to zero and the equation separates into three independent wave equations for each component of $\bm{E}$, i.e., it is sufficient to consider each component as a scalar field (scalar diffraction theory).
This approximation is valid for a dielectric haloscope with sufficiently homogeneous disks and has explicitly been confirmed for the ideal system discussed in the next section, as we show explicitly in appendix~\ref{app:comparision}.
However, the calculations below can also be easily generalized by solving for the modes of the vectorized equation, see e.g.~\cite{yeh2008essence,snyder1983optical}.

\begin{figure}
    \centering
    \includegraphics[width=0.63\textwidth]{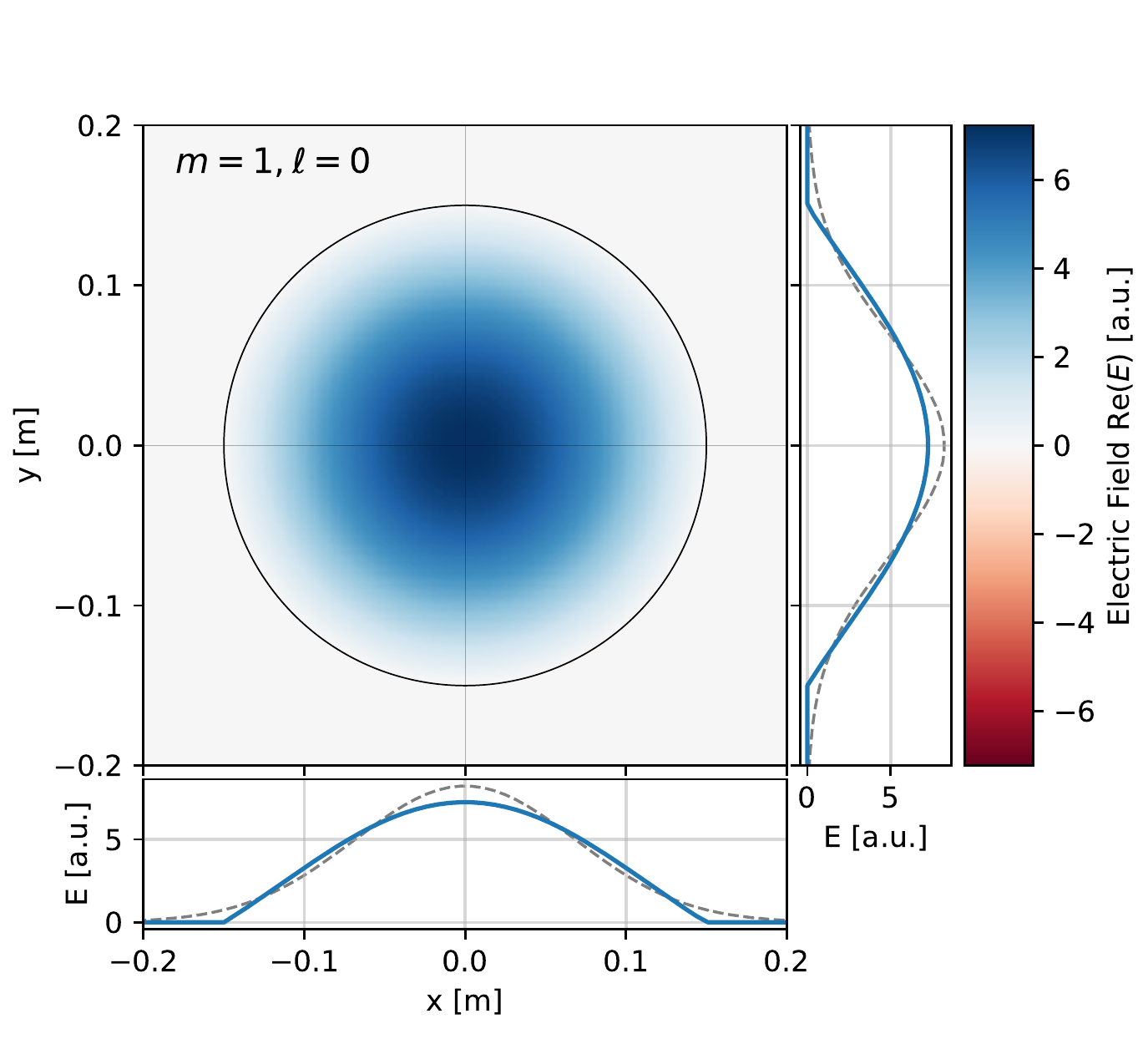}
    \includegraphics[width=0.315\textwidth]{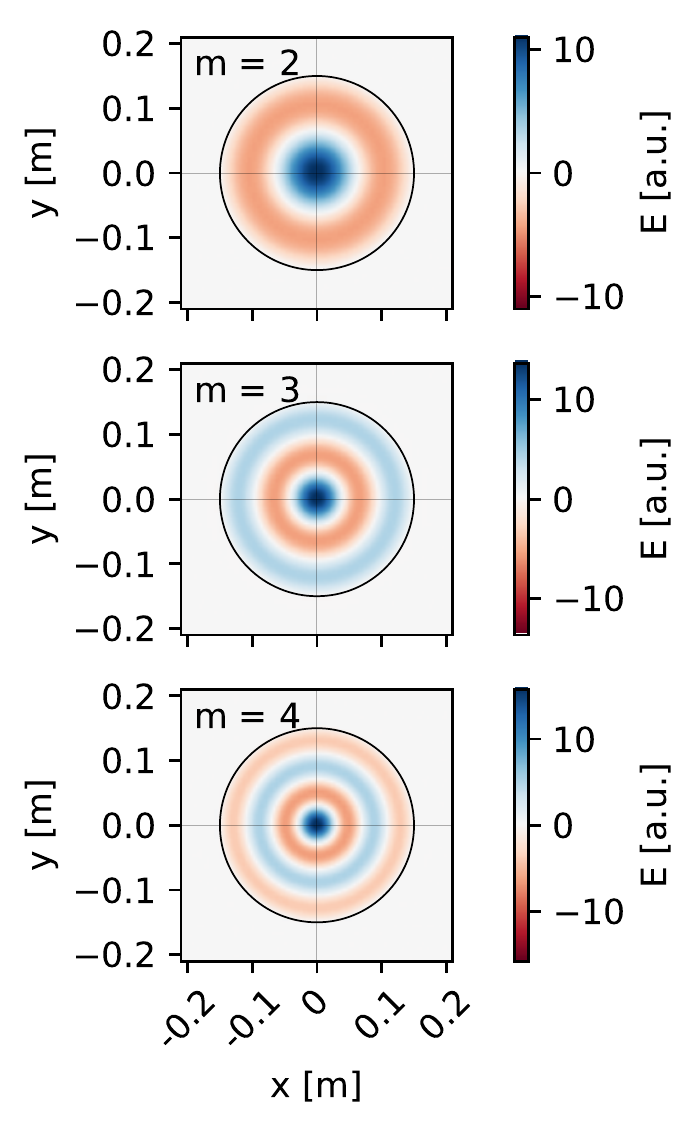}
    \caption{Modes of the circular dielectric haloscope. \textbf{Left:} Fundamental mode $m = 1, \ell = 0$. Main panel: Spatial field distribution. Side and bottom panels: Fields on the $x$ and $y$ axis (at~$y=0,x=0$, respectively) (blue solid lines) and fields corresponding to a Gaussian beam~\cite{goldsmith1998quasioptical} with a beam waist radius of $w_0 \approx \o/3$ (gray dashed lines), discussed later in section~\ref{sec:ideal_system}. \textbf{Right:} Same as main panel on the left, but for higher modes with $\ell = 0$.}
    \label{fig:mode-patterns}
\end{figure}{}

To begin with, consider a cylinder of dielectric material with radius $R$ (diameter $\o = 2R$) surrounded by vacuum forming a dielectric waveguide.
In the limit of large radius $R \gg \lambda$ and large dielectric constant $\epsilon \gg 1$, we obtain a model for one of the disks of the dielectric haloscope.
In this limit the electric fields drop to zero at the outer boundary $r=R$ of the disk. 
Explicitly, the solutions to the source-free scalar wave equations (i.e., eq.~\eqref{eq:sim:3d:waveequation-td} with $\nabla \cdot \bm{E} = 0$ and $a = 0$) are the eigenmodes which are illustrated in figure~\ref{fig:mode-patterns} and given by~\cite{snyder1983optical,yeh2008essence}
\begin{equation}
    E_{m\ell}(r,\phi) = \mathcal{N}_{m\ell} J_\ell(k_{c,m\ell} r) e^{i \ell \phi} \quad , \quad  J_\ell(k_{c,m\ell} R) = 0,
    \label{eq:modes}
\end{equation} 
with discrete radial mode indices $m > 0$ and azimuthal mode indices $\ell$. $J_\ell$ is the Bessel function of the first kind of order~$\ell$, and we take $\mathcal{N}_{m\ell}$ as a normalization factor such that $\int |E_{m\ell}|^2 dA = 1$.
Here, $k_c$ is the transverse momentum, i.e., the momentum in the disk plane.
These modes are orthogonal and complete in the sense that we can expand any field distribution inside of the disks into a set of these modes. Most importantly, they propagate independently along the $z$-direction within the~disks~as
\begin{equation}
    E(r,\phi,z) = \sum_{m,\ell} e_{m\ell} E_{m\ell}(r,\phi)  e^{-i k_{z,m\ell} z} \quad , \quad   k_{z,m\ell} = \sqrt{k_0^2 - k_{c,m\ell}^2},
    \label{eq:modes-wgprop}
\end{equation}
where $e_{m\ell}$ are the coefficients for the mode expansion, $k_{z,m\ell}$ is the propagation constant and $k_0 = \sqrt{\epsilon} \omega$.
In free space these eigenmodes of the dielectric disks in general do not propagate independently anymore, because they are no longer solutions of the scalar wave equation under the free space boundary conditions. Since they are orthogonal and complete, we still can expand fields at $r < R$ into these modes, but during propagation they mix with each other,~i.e.,
\begin{equation}
    E(r,\phi,z) = \sum_{m,\ell,m',\ell'} P^{m' \ell'}_{m\ell}(z)~e_{m' \ell'}~E_{m\ell}(r,\phi),
    \label{eq:modes-fsprop}
\end{equation}
with the linear map $P^{m' \ell'}_{m\ell}(z)$ between the modes. $P$ can be calculated by using the scalar diffraction theory in free space discussed in~\cite{Knirck:2019eug}. $P^{m' \ell'}_{m\ell}$ is the coefficient of the mode $(m',\ell')$ when expanding the field obtained after propagating the mode $(m,\ell)$ for a distance $z$ in free space.
One can generalize the 1D transfer matrix formalism for dielectric haloscopes in~\cite{millar2017dielectric} by having left-moving and right-moving fields for each mode in each region and by directly including the mixing matrix $P$, see e.g.~\cite{Knirck:2020}.
We refer to this kind of calculation as~Mode~Matching, because on an interface between two media with different dielectric constants the sum of modes describing the fields on one side needs to be matched with the respective sum of modes on the other side.

\begin{table}[]
    \centering
    \begin{tabular}{cccccccc} 
		\toprule
		& Axion Coupling & & \multicolumn{5}{c}{Propagation}\\ 
		& & & \multicolumn{2}{c}{$\o = \SI{30}{\centi\metre}$}&& \multicolumn{2}{c}{$\o = \SI{1}{\metre}$} \\
		Mode ($\ell = 0$) & $|\eta_{m\ell}|^2$  
		&& $k_{c,m\ell}\,[{\rm m}^{-1}]$ & $\delta_{{\rm d},m\ell}$&& $k_{c,m\ell}\,[{\rm m}^{-1}]$ & $\delta_{{\rm d},m\ell}$ \\ 
		\midrule 
		$m=1$ & \SI{69}{\percent} 
		&& 16 & $\num{2e-5}$ && 5 & $\num{5e-7}$ \\ 
		$m=2$ &\SI{13}{\percent}  
		&& 37 & $\num{1e-4}$ && 11 & $\num{2e-6}$\\ 
		$m=3$ &\SI{5}{\percent}  
		&& 58 & $\num{2e-4}$ && 17 & $\num{6e-6}$ \\ 
		$m=4$ &\SI{3}{\percent}  
		&& 79 & $\num{5e-4}$ && 23 & $\num{1e-5}$ \\ 
		$m=5$ &\SI{2}{\percent}  
		&& 100 & $\num{7e-4}$ && 30 & $\num{2e-5}$ \\ 
		$m=6$ &\SI{1}{\percent}  
		&& 121 & $\num{1e-3}$ && 36 & $\num{3e-5}$\\ 
		\multicolumn{8}{c}{$\vdots$} \\ 
		\bottomrule
	\end{tabular}
    \caption{Properties of the most important modes for a dielectric haloscope, as envisioned for the MADMAX prototype (disk diameter $\o = \SI{30}{\centi\metre}$) and final scale experiment (disk diameter $\o = \SI{1}{\metre}$). $|\eta_{m\ell}|^2$ denotes the coupling of the uniform axion field under a uniform magnetic field to the mode, $k_{c,m\ell}$ its transverse momentum and $\delta_{{\rm d},m\ell}$ the diffraction loss parameter between the dielectric disks as defined in the text at \SI{22}{\giga\hertz}.}
    \label{tab:mode-properties}
\end{table}
In order to see which modes are relevant for a dielectric haloscope, we have to consider their coupling to the axion-induced field $E_a$. Table~\ref{tab:mode-properties} summarizes the most important modes for the dielectric haloscope for a uniform external magnetic field and negligible axion velocity.
The coefficients $\eta_{m\ell}$ refer to the coupling of the mode to the axion-induced field~$E_a$, i.e., they are the coefficients of the modes when expanding $E_a$ on the disk surfaces into the modes. Explicitly,
\begin{equation}
    \eta_{m\ell}(z) = \tfrac{1}{\mathcal{N}_a} \int E_{m\ell}^*(r,\phi) \cdot E_a(r,\phi,z) \, d A = \tfrac{1}{\mathcal{N}_a'}  \int E_{m\ell}^*(r,\phi) \cdot \epsilon^{-1} \bm{B}_{\rm e}(r,\phi,z) \, a(r,\phi,z) \, d A,
    \label{eq:modes-fsprop}
\end{equation}
with a normalization factor $\mathcal{N}^{(')}_a$ such that $\sum |\eta_{m\ell}|^2 = 1$. 
For our axion haloscope actually only the azimuthally symmetric ($\ell = 0$) lower modes with $m = 1,2,3,4$ have a coupling stronger than $2\%$ to the axion-induced electric field. All modes with $\ell \neq 0$ do not couple due to symmetry, although imperfections may affect $\eta$, see section~\ref{sec:nonideal_system}.
When only considering these relevant modes with $m = 1,2,3,4; \ell = 0$ even for disks with diameter $\o = \SI{30}{\centi\meter}$ at \SI{22}{\giga\hertz} the mixing between the modes, i.e., $|P^{m'\ell'}_{m\ell}|$ for $(m, \ell) \neq (m', \ell')$, is smaller than $\approx \num{8e-4}$.
So unless the system is tuned to be very resonant, the mixing can be neglected, i.e., $P^{m'\ell'}_{m\ell}$ becomes diagonal and can be written as
\begin{equation}
{P^{m'\ell'}_{m\ell} \approx \exp{(i k_{z, m\ell} z - \tfrac{1}{2} \delta_{{\rm d}, m\ell} k_{z, m\ell} z)}\, \mathbb{1}^{m'\ell'}_{m\ell}}, \label{eq:sim:3d:tools:modes:mix-independent}
\end{equation}
where $\delta_{{\rm d},m\ell}$ is a diffraction loss parameter and all modes propagate essentially independently. The parameter $\delta_{{\rm d},m\ell}$ can be suppressed by using disks with larger diameters, as expected.
If,~on the other hand, the system is tuned to be very resonant for a specific mode, the difference in $k_{z,m\ell}$ for the other modes will make them rapidly dephase, i.e., make all other modes irrelevant.

\section{Ideal 3D Booster}
\label{sec:ideal_system}
We first consider an ideal but 3D booster with disks of finite extent, which are however still perfectly flat and parallel. We study two benchmark systems, tuned to an axion mass of $m_a \approx \SI{90}{\micro\electronvolt}$ ($\nu \approx \SI{22}{\giga\hertz}$).
The optimal boost factor bandwidth is given by a trade-off between disk readjustment time for tuning, and actual data taking time. The minimum bandwidth is further limited by losses. Here we consider a bandwidth of $\approx \SI{50}{\mega\hertz}$ close to preliminary estimates of the optimal bandwidth maximizing scan speed for MADMAX~\cite{Brun:2019lyf}.
We consider a booster with 20 lanthanum aluminate disks (assuming an isotropic dielectric constant of $\epsilon = 24$) with a disk diameter of $\o = \SI{30}{\centi\meter}$ and thickness \SI{1}{\milli\metre} as presently foreseen for the MADMAX prototype; in addition, we examine an 80 disk system with a disk diameter of $\o = \SI{1}{\meter}$ as envisioned in the final MADMAX setup~\cite{Brun:2019lyf,Beurthey:2020yuq}.
All presented simulations assume free space surrounding for simplicity, see also~\cite{Knirck:2019eug,Bergermann:2019}.
This setup is expected to maximize diffraction losses. 
Detailed studies on the impact of using other different boundary conditions, e.g., conducting walls, will be discussed in future works.

\begin{figure}
    \centering
    \includegraphics[width=\textwidth]{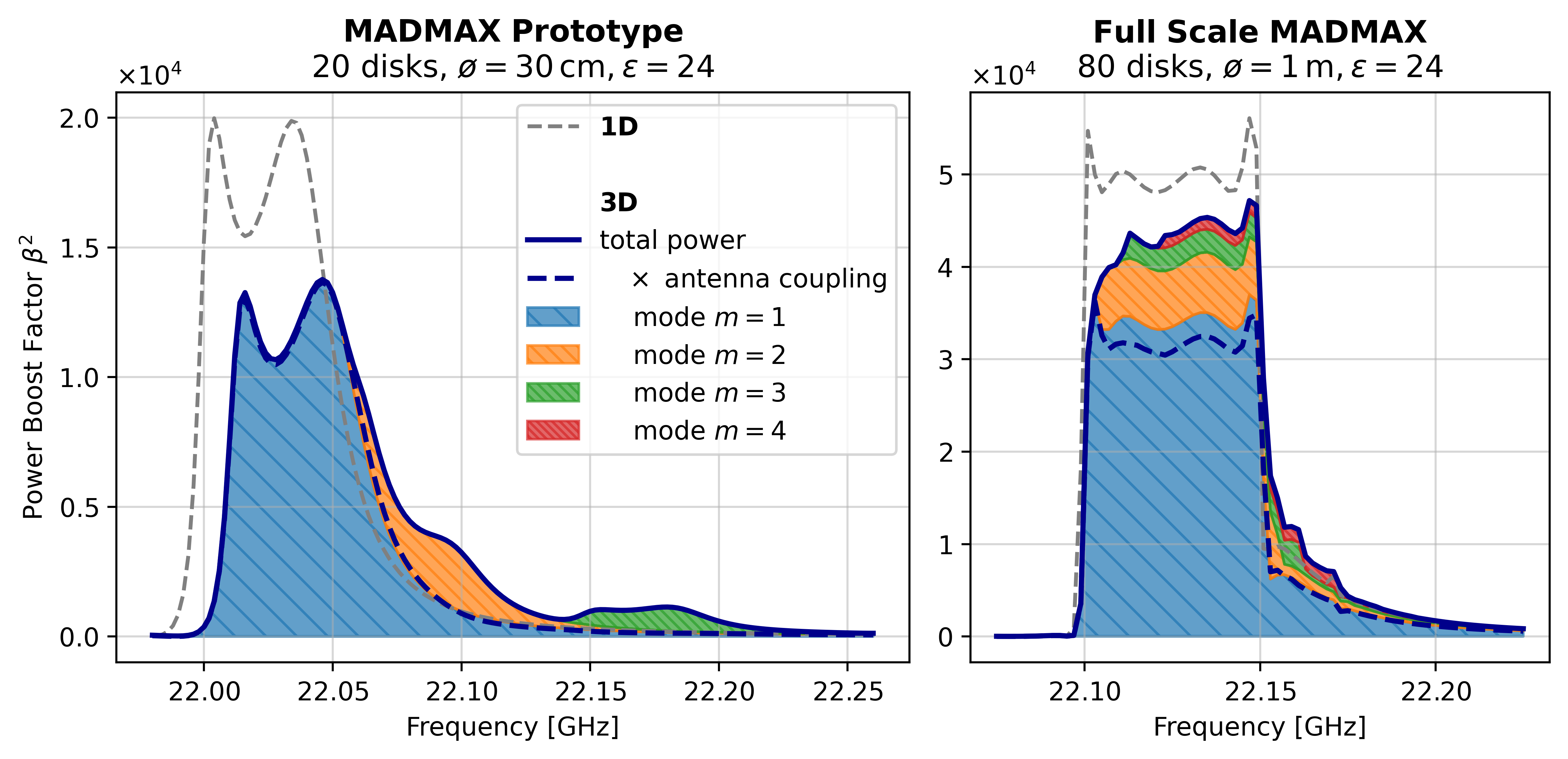}
    \caption{Power boost factor $\beta^2$ considering dielectric disks of finite size with spacings tuned to cover a bandwidth of $\approx \SI{50}{\mega\hertz}$ at around $\SI{22}{\giga\hertz}$ ($m_a \approx \SI{90}{\micro\electronvolt} $), in terms of the 1D analytical result following~\cite{millar2017dielectric} (gray dashed line), total emitted power in 3D (solid blue line), the power which can be coupled to a Gaussian beam antenna (dashed blue line) and total emitted power contributed by different modes (differently colored hatched regions). \textbf{Left:} For 20 disks with a diameter of \SI{30}{\centi\meter} as in the proposed MADMAX prototype (antenna $w_0 = \SI{10}{\centi\meter}$). \textbf{Right:}~For 80 disks with a diameter of \SI{1}{\meter} as in the proposed MADMAX final experimental phase (antenna $w_0 = \SI{30}{\centi\meter}$). Consistently obtained with different numerical methods (2D3D~FEM, Recursive Fourier and Mode Matching), see appendix~\ref{app:comparision}.
    }
    \label{fig:ideal}
\end{figure}

Figure~\ref{fig:ideal} shows the power boost factor of such systems in terms of total emitted power (solid blue) and the power which can be coupled to an antenna receiving Gaussian beams as defined in~\cite{goldsmith1998quasioptical} with beam waist radius $w_0 \approx \o/3$ (dashed blue) compared to the 1D result (dashed gray). 
The double-peak or four-peak substructure, respectively, corresponds to different contributing resonances, for more details see~\cite{millar2017dielectric}.
Results from different numerical methods, i.e., 2D3D~FEM, Recursive Fourier Propagation and Mode Matching are consistent up to percent level, 
which is negligible for the experiment's sensitivity to axion CDM and the axion-photon coupling $|C_{a\gamma}|$.
This confirms the validity of the scalar diffraction theory for the idealized system, since the 2D3D~FEM method directly solves the full vectorized wave equation~\eqref{eq:sim:3d:waveequation-td}. For more details see appendix~\ref{app:comparision}.

Turning our attention to the results themselves, we first notice that the boost factor curve is shifted to higher frequencies compared to the 1D calculation. This is easily understood considering the phase evolution of the different modes along the booster. Due to the transverse momentum $k_c$ of the modes the phase changes slower along the $z$-direction compared to the 1D case according to eq.~\eqref{eq:modes-wgprop}. Therefore, in order to have the same resonant behavior as in 1D one needs to ``speed up'' the phase evolution by going to slightly higher frequencies. 
For the lower modes $(m,\ell)$ with small transverse momenta $k_c \ll k_z$ the frequency shift compared to the 1D calculation is
\begin{equation}
\Delta \nu_{m\ell} \approx \frac{1}{8 \pi^2} \frac{k_c^2}{\nu} \approx \SI{13}{\mega\hertz} \left(\frac{j_{\ell,m}}{j_{0,1}}\right)^2 \left(\frac{\SI{30}{\centi\metre}}{\o}\right)^2 \left(\frac{\SI{22}{\giga\hertz}}{\nu}\right),
\label{eq:sim:freqshift}
\end{equation}
where $j_{\ell,m}$ is the $m$-th zero of $J_\ell$, which roughly scales linearly with $m$. 
Since higher modes have higher transverse momenta, cf.\ table~\ref{tab:mode-properties}, the shift is more pronounced for higher modes. 
As each mode propagates essentially independently through the system,
no matter how the disk spacings in the system are tuned, for a fixed disk diameter the different modes always appear at the same frequency shifts relative to each other.
The bandwidth above which higher modes start to become relevant is therefore $\Delta \nu_{\beta} \approx \Delta \nu_{2\,0} - \Delta \nu_{1\,0}$, which gives \SI{55}{\mega\hertz} for the prototype booster and \SI{5}{\mega\hertz} for the full-scale booster, consistent with figure~\ref{fig:ideal}.

Now considering the power emitted by the system, we see that in 3D the boost factor is reduced compared to the 1D calculations.
Since all modes are orthogonal, the total power emitted is simply the sum of the power carried by each mode, as indicated by the stacked hatched regions in figure~\ref{fig:ideal}.
In the benchmark case for the MADMAX prototype (left) we see that the second mode is already shifted by almost the full bandwidth of the boost factor itself and we essentially only get the power contributed by the first mode within a \SI{50}{\mega\hertz} bandwidth. Since this mode couples to $\SI{69}{\percent}$ (independent of disk diameter) to the axion field, the boost factor is reduced by up to $\SI{30}{\percent}$ compared to the 1D case.
This effect should be seen as a reduced coupling efficiency (form factor) of the system to the axion field and not as (diffraction) loss. Indeed, the diffraction loss of the first mode arising from the finite disk size in this case is smaller than $\delta_{\rm d} \approx 10^{-5}$ at this frequency (see table~\ref{tab:mode-properties}) which is negligible. This may not hold anymore when we consider the geometrical inaccuracies in section~\ref{sec:nonideal_system}.

Lastly, we have to consider how to couple the power leaving the booster with an antenna into a receiver. The fundamental mode has a frequency-independent $97\%$ matching ratio  with a Gaussian beam~\cite{goldsmith1998quasioptical} with a beam waist radius of $w_0 \approx \o/3$, see also figure~\ref{fig:mode-patterns}\,(left). We therefore consider the coupling efficiencies to Gaussian beam antennas with this beam shape in this paper. Hence, in case only the fundamental mode contributes,
we can achieve very good coupling efficiencies.
In case the total power is also carried by higher modes, like in the 80 disk calculation in figure~\ref{fig:ideal}\,(right), one can only receive significant power provided by the fundamental mode with the Gaussian antenna.
This contribution is still $\gtrsim \SI{70}{\percent}$ of the total power due to the coupling of the axion field to the fundamental mode.
However, small couplings of the higher modes to the Gaussian may interfere destructively when coupled to the antenna, further decreasing the received power. In principle it is possible to design an antenna which is matched to a more optimal combination of modes, as long as their relative phase stays roughly constant over the boost factor bandwidth -- or in other words the total beam shape does not change drastically with frequency. For the initial stage of dielectric haloscopes this may already be a too elaborate approach. 
In summary, as long as the boost factor bandwidth is smaller than the difference between the frequency shifts of the first two modes, the optimal antenna is one that couples only to the fundamental mode.
In particular, for the MADMAX prototype, designed for the frequency range from $18$ to \SI{24}{\giga\hertz}, an antenna system which couples to a Gaussian beam with beam waist radius of approximately \SI{10}{\centi\metre} is close to optimal.

Not considered here, but crucial for a final experimental realization, might be possible reflections on the antenna, especially those of the higher modes, which after the reflection may couple and interfere destructively with the fundamental mode. For MADMAX such reflection effects have been already experimentally studied in~\cite{Egge:2020hyo}. There it was demonstrated on a 5-disk setup that adverse effects due to reflections may be significantly reduced by absorbing unwanted radiation in the vicinity of the antenna and calibrating out residual reflections using a dedicated model.

\section{Non-Ideal Effects}
\label{sec:nonideal_system}
A realistic system will always have inaccuracies, contrary to what was assumed in the previous section.
Therefore, in the following we study the influence of axion velocity effects and inhomogeneities of the external magnetic field (causing changes to the axion-induced field~$E_a$), as well as geometrical imperfections (tilts, planarity, surface roughness and tiling of the disks).

\subsection{Axion Velocity}
\label{sec:nonideal_system:velocity}

With non-zero axion velocity $\bm{v}_a$ the axion field $a$ and therefore also the axion-induced electric field $E_a$ acquire a spatial phase factor $\exp(-i m_a \bm{v}_a \bm{x})$ over the setup. A velocity along the booster axis causes phase differences between the disks and has been studied already in~\cite{Millar:2017eoc,Knirck:2018knd}. 
A transverse velocity $v_{a,\parallel}$ 
tilts the otherwise perpendicular angle of emission
from the individual disks~\cite{Jaeckel:2013sqa,Jaeckel:2015kea,Jaeckel:2017sjb}. 
We can study the effect of this tilting by decomposing the axion-induced field into the above modes and observing how the coupling efficiencies $\eta_{m\ell}$ change with transverse velocity.
For the fundamental mode one finds analytically
\begin{equation}
    \frac{\Delta |\eta_{10}|^2}{|\eta_{10}|^2} \approx \frac{j_{0,1}^2 - 4}{2\,j_{0,1}^2} \left(m_a v_{a,\parallel} R \right)^2 \approx \SI{1}{\percent} \left( \frac{m_a}{\SI{100}{\micro\electronvolt}}\right)^2 \left(\frac{v_{a,\parallel}}{10^{-3}c} \right)^2 \left( \frac{\o}{\SI{1}{\metre}} \right)^2.
\end{equation}
This holds for small axion velocities, i.e., $m_a v_{a,\parallel} R \ll 1$, which is applicable for the MADMAX boosters below about $m_a < \SI{500}{\micro\electronvolt}$. An exact result can be found in appendix~\ref{app:velocity}.
Below $m_a=\SI{100}{\micro\electronvolt}$ the effect on the full-scale MADMAX sensitivity is negligible.
Although the effects may become more relevant for higher masses of $m_a=100 - \SI{400}{\micro\electronvolt}$, still $\gtrsim \SI{90}{\percent}$ of power is left in the fundamental mode. In particular, one would have to take the average of the signal power over the CDM velocity distribution rather than just considering one velocity. 
Typical data-acquisition times for MADMAX before tuning to the next frequency band are expected to be at the order of a few days~\cite{Brun:2019lyf}. Since the earth rotates within the CDM `wind', some of the velocity effects will average out and make the above reduction even milder.

\begin{figure}
    \centering
    \includegraphics[width=\textwidth]{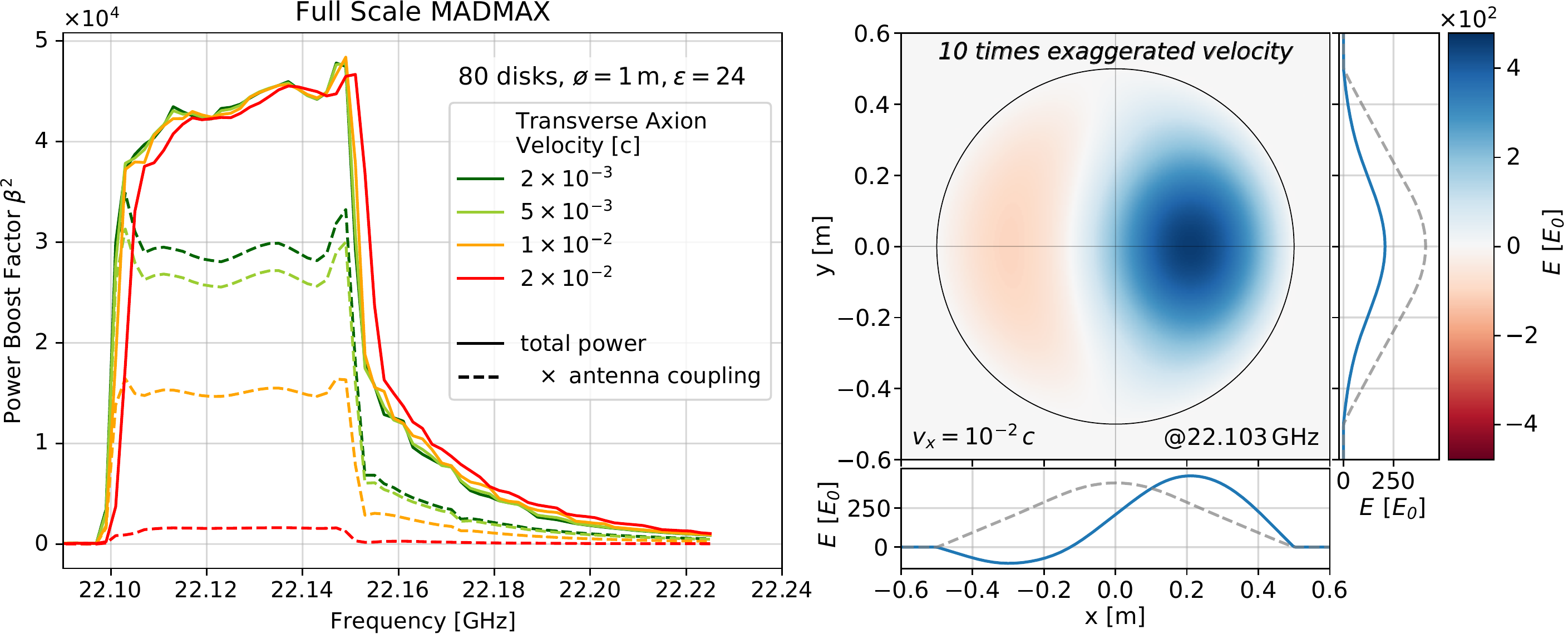}
    \caption{Effect of transverse axion velocity on the benchmark boost factor for the full-scale MADMAX booster. \textbf{Left:} Power boost factor $\beta^2$ for different fixed transverse axion velocities (different colors). The boost factor corresponding to the total emitted power is denoted by the solid lines while the amount which can be coupled to a $w_0 = 30\,{\rm cm}$ Gaussian beam antenna is indicated by the dashed lines. \textbf{Right:} Beam shape in terms of the electric field of the emitted wave from the booster at a frequency of \SI{22.103}{\giga\hertz} for an exaggerated transverse axion velocity of $v_x \approx  10^{-2}c$ (around $10$ times larger than typical CDM velocities), in order to make the effect on the beam shape visible. The side panels show the field at the $x$ and $y$ axes (at $y=0,x=0$, respectively) as a blue line and the field when setting the velocity to zero as a dashed gray line. We show the fields at a fixed instant of time.}
    \label{fig:velocity}
\end{figure}

The effect on the boost factor is explicitly demonstrated on the benchmark boost factor for the full-scale MADMAX booster in figure~\ref{fig:velocity}\,(left) for an axion velocity exaggerated by up to one order of magnitude compared to realistic CDM velocities of $v_a\approx 10^{-3}c$. 
We see that the received power (dashed lines) is degraded while the total power emitted by the haloscope (solid lines, almost on top of each other) remains almost unchanged, but is in the modes that do not couple to the antenna.
No curve is shown for $v_a = 10^{-3}c$, since already $v_a = \num{2e-3} c$ does not significantly change the boost factor compared to the zero velocity case.
For even higher axion velocities (not shown) nearly all power is contained in higher modes. Since they have higher $k_c$ the total power boost factor shifts to higher frequencies. Higher modes are more prone to diffraction losses and the inaccuracies of the setup described in the following sections. Therefore, also the total power emitted is reduced for higher velocities. 
For realistic CDM velocities of $v_a\approx 10^{-3}c$, however, our benchmark boost factors are not changed significantly.

A finite axion velocity slightly tilts the emissions from individual disks. Therefore, the center of the beam emitted from the booster shifts away from the center of the disk as shown in figure~\ref{fig:velocity}\,(right).
This effect could in principle be used to build a velocity-sensitive haloscope after the discovery of the axion and to investigate and measure properties of the local dark matter halo, see for example~\cite{Jaeckel:2013sqa, Knirck:2018knd}.

\subsection{Magnetic Field Inhomogeneity}
Analogous to the velocity effects above, a transverse inhomogeneity of the magnetic field implies a corresponding inhomogeneity in the axion-induced field $E_a$. Therefore, it changes the amount of power coupled into the different modes. For example a magnetic field proportional to the beam shape of the fundamental mode would cause the coupling efficiency of the first mode to be $|\eta_{10}|^2 = \SI{100}{\percent}$. 
Realization of such a magnet is, however, technically challenging and may increase magnet cost~significantly.
Here we consider a magnetic field amplitude with azimuthal and radial inhomogeneity, which is motivated by the symmetry of typically considered dipole magnets. We consider such a magnetic field parametrized by
\begin{equation}
    {\bm B}_{\rm e}(r, \phi) = B_0 \left[1 + h \sin(k \phi) \frac{r^2}{R^2} \right] \hat{\bm e}_y,
\end{equation}
where $B_0$ is the magnetic field amplitude, $h$ is the maximum relative scale of the inhomogeneity on the disk, $R = \o/2$ and $k$ is a non-zero positive integer.
For small $h \ll 1$ one can show that the relative change in the coupling coefficients of the $\ell = 0$ modes happens only at second order in $h$ and is given by
\begin{equation}
    \frac{\Delta |\eta_{m0}|^2}{|\eta_{m0}|^2} \approx -\frac{1}{6} h^2,
\end{equation}
i.e., radial symmetric transverse inhomogeneities at the \SI{10}{\percent} level leave the mode coupling coefficients unchanged well below the percent level. Therefore, such inhomogeneities have insignificant impact on sensitivity.
This result has been confirmed with explicit numerical calculations using Recursive Fourier Propagation.

\subsection{Geometrical Inaccuracies of the Dielectric Disks}
\label{sec:nonideal_system:thicknessvar}
Next we consider geometrical inaccuracies such as disk tilts, disk planarity and surface roughness.
These mainly affect propagation of electromagnetic waves within the booster. If the distance between two interfaces varies as $\Delta z(r,\phi)$ in the plane parallel to the disk surfaces (transverse thickness variation), the corresponding phase changes during propagation give rise to additional mode mixing as
\begin{equation}
P^{m'\ell'}_{m\ell} = \int E_{m\ell}^*(r,\phi) E_{m'\ell'}(r,\phi) \exp\left[i k_0 \Delta z(r,\phi)\right] \, {\rm d} A, \label{eq:sim:3d:results:tilt:mixing}
\end{equation}
where we have left out the propagation and corresponding diffraction by a distance $z$ in this formula for clarity (it is included in the simulations below).
The phase factor is most relevant at places where both $E_{m\ell}^*(r,\phi)$ and $E_{m'\ell'}(r,\phi)$ are maximized. Therefore, inaccuracies in the center of the disks are in general most relevant.

\begin{figure}
	\centering
	\includegraphics[width=\textwidth]{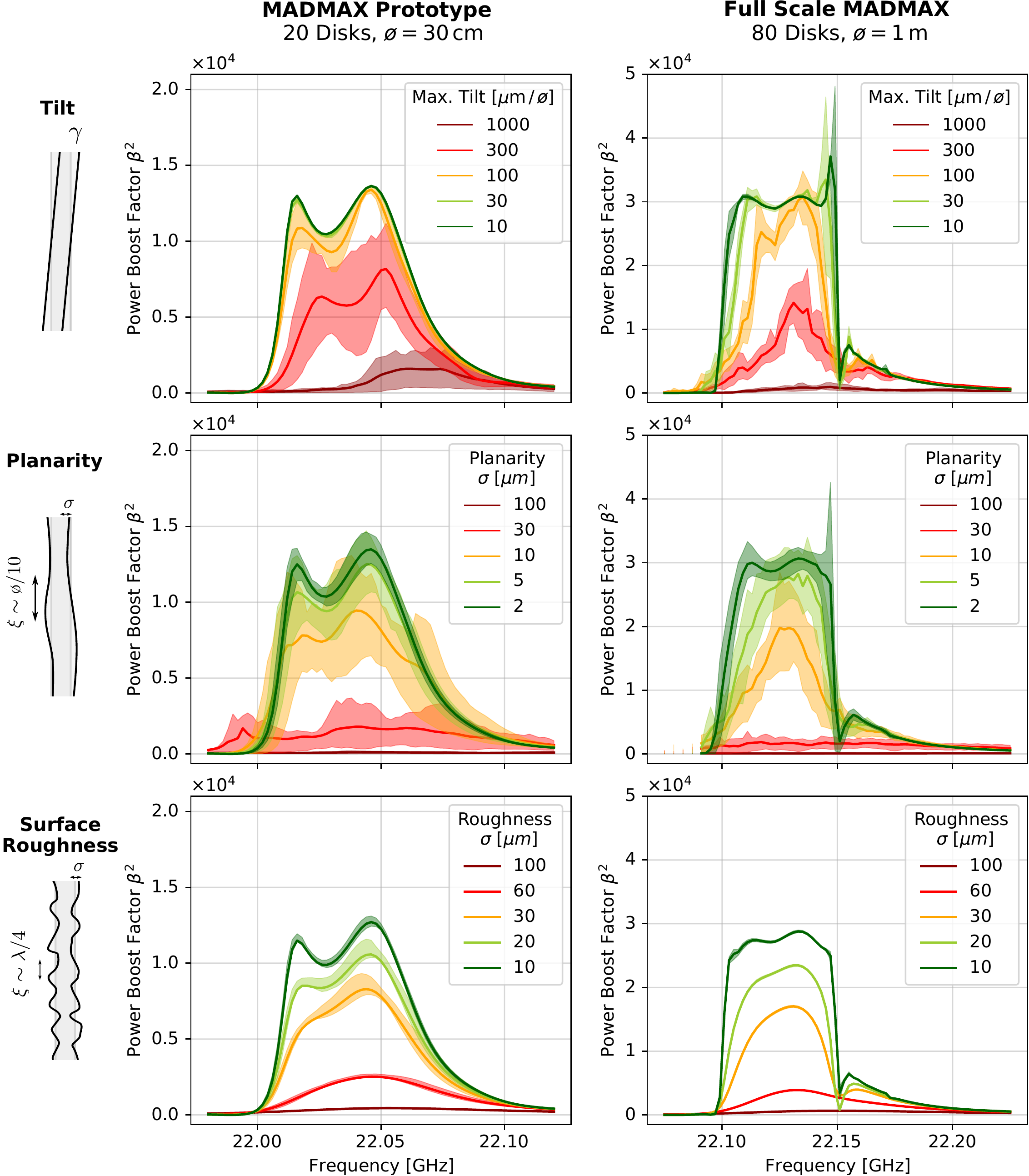}
	\caption{Effect of different geometrical disk inaccuracies on the benchmark power boost factor times antenna coupling assuming a Gaussian beam with $w_0 \approx \o/3$.  The left column shows the results for a 20 \LALO{} disk booster with \SI{30}{\centi\metre} disk diameter, the right column shows the results for a 80 \LALO{} disk booster with \SI{1}{\metre} diameter. The top row shows randomly tilted disks, where the maximum tilt in both $x$ and $y$ direction of each disk is as indicated in the legends. The middle row shows non-planar disks (thickness variations with correlation length of $\xi \approx \o / 10$) of scale $\sigma$ as indicated in the legends. The bottom row shows the same but for a correlation length of $\xi \approx \lambda / 4$ (surface roughness). 
	The solid lines each refer to the ensemble mean and the shaded regions each to the range between the \SI{16}{\percent} to \SI{84}{\percent} percentiles ($1\sigma$ for a Gaussian distribution). Different colors refer to different magnitudes of distortion as specified in the respective panel legend.}
	\label{fig:sim:3d:result:roughness-bf}
\end{figure}

We parametrize $\Delta z$  as a random function where $\sigma$ is the root-mean-square of the elevation and $\xi$ the transverse correlation length, i.e., the standard deviation characterizing the radius of a typical bump, cf.~figure~\ref{fig:sim:3d:result:roughness-bf}.
For large $\xi \gg k_{c,m\ell}^{-1}$ we can approximate
\begin{equation}
	P^{m'\ell'}_{m\ell} \approx \exp(i k_0 \langle\Delta z\rangle) ~ (P_0)^{m'\ell'}_{m\ell}, \label{eq:sim:3d:results:tilt:mixing-phase-out}
	\end{equation}
where $\langle ... \rangle$ denotes the average and $P_0$ is the mixing matrix for the unperturbed system. Thus in the limit of large $\xi$ there is only an overall phase error analogous to a misplacement of the disks.
On the other hand, for small $\xi \ll k_{c,m\ell}^{-1}$, one finds
\begin{equation}
	P^{m'\ell'}_{m\ell} \approx (1- \tfrac{1}{2} k_0^2 \sigma^2) ~ (P_0)^{m'\ell'}_{m\ell}, \label{eq:sim:3d:results:tilt:mixing-loss} 
\end{equation}
where the system is dominated by an effective loss, while the phase errors are averaged~out.
This effective loss can be parameterized within the disks as (analogously to the definition of $\delta_{\rm d}$ in eq.~\eqref{eq:sim:3d:tools:modes:mix-independent})
\begin{equation}
	\delta_{\Delta z} \approx 2 \pi \left(\frac{\sqrt{\epsilon} \sigma^2}{\lambda d}\right) \approx \num{2e-3} \left(\frac{\sigma}{\SI{30}{\micro\metre}}\right)^2 \left(\frac{\SI{1}{\milli\metre}}{d}\right) \left(\frac{\epsilon}{24}\right)^{\frac{1}{2}}  \left(\frac{\nu}{\SI{22}{\giga\hertz}}\right),
\end{equation}
with the disk thickness $d$. The equation holds analogously within the free space gaps, but there it gives around two orders of magnitude smaller $\delta_{\Delta z}$ (because of $\epsilon=1$ and $d \approx \si{\centi\metre}$ there).
The effects in both of these limits for $\xi$ can be estimated using 1D calculations as in~\cite{millar2017dielectric}.
For the intermediate range where $\xi \approx k_{c,m\ell}^{-1}$ both phase errors and effective loss are relevant. In addition, $P$ will not be well approximated by a diagonal matrix anymore (`mode mixing'), which gives the strongest design constraints, as we will see below. 
We evaluate representative elements of the mixing matrix for different correlation lengths explicitly in appendix~\ref{app:mxing_thicknessvar}.

We extend these estimates with explicit numerical results shown in figure~\ref{fig:sim:3d:result:roughness-bf}. We survey the effect on the benchmark power boost factor for both the MADMAX prototype (left column) and full-scale MADMAX (right column).
To this end we consider uniformly distributed random tilts~$\gamma$ of the dielectric disks around both $x$ and $y$ axis through their center (first row), non-planar disks ($\xi = \o/10$, second row) and surface roughness ($\xi \approx \lambda/4$, third row).
For each case we take many random samples of a respectively deformed booster and calculate the boost factor times antenna coupling for a Gaussian beam antenna as discussed above. 

Calculations with 20 disks are feasible with both Recursive Fourier Propagation and the Mode Matching methods and lead to consistent results, cf.~also appendix~\ref{app:comparision-non-ideal}.
For surface roughness Mode Matching gives more conservative results.
80 disk calculations have been only feasible with the Mode Matching method, as too many iterations are needed to achieve convergence with Recursive Fourier Propagation.

In order to leave the power boost factor $\beta^2$ unchanged on the level of $\lesssim 20\,\%$ for each individual effect considered alone, we conclude that tilts at the order of $\gamma \lesssim \SI{100}{\micro\metre} / \o$ are required, planarity on length scales of $\xi \approx \o/10$ should be $\sigma \lesssim \SI{5}{\micro\metre}$ and surface roughness is allowed to be $\sigma \lesssim \SI{20}{\micro\metre}$.
In an engineering context often the deviation between the minimum and maximum (min-to-max) $\Delta z_{\rm min-max}$ is quoted instead of $\sigma$. We note that for planarity $\Delta z_{\rm min-max} \approx 4\sigma$, and surface roughness $\Delta z_{\rm min-max} \approx 6\sigma$.

By defining the tilt around a central axis we have $\langle \Delta z \rangle = 0$ and thus suppressed phase errors for the tilts here.
This separates the requirement on the tilt from the overall position accuracy of the dielectric disks,
which gives more stringent constraints ($\lesssim \SI{5}{\micro\metre}$ in this case~\cite{Millar:2017eoc,Knirck:2020}). 
The strongest design constraints in this section arise for planarity. This is intuitive, since the considered transverse thickness variations $\Delta z$ appear on similar length scales $\xi$ as the most relevant modes, maximizing mode mixing effects.
The results for surface roughness are consistent with 1D calculations taking losses at the order estimated in eq.~\eqref{eq:sim:3d:results:tilt:mixing-loss} into account. 
These constraints remain approximately unchanged when increasing the number of dielectric disks from 20 to 80 disks but keeping the desired boost factor bandwidth the same. This is expected, since a bandwidth of \SI{50}{\mega\hertz} naively corresponds to a resonance with the beam experiencing about $\SI{20}{\giga\hertz} / \SI{50}{\mega\hertz} \approx 400$ bounces before leaving the booster independently of how many disks are actually installed.

These results are expected to generalize to boost factors at other frequencies with the same relative boost factor bandwidth ($\nu / \Delta \nu \approx 400$) when written in units of the wavelength~$\lambda$.
This can for example be seen from eq.~\eqref{eq:sim:3d:results:tilt:mixing}, which gives the same $P$ at different frequencies when scaling $\Delta z$ accordingly.
The above constraints then read: maximum tilts at the order of $\gamma \lesssim \num{7e-3} \lambda / \o$, planarity of $\sigma \lesssim \num{4e-4} \lambda$ (on length scales of $\xi \approx \o/10$) and maximum surface roughness of $\sigma \lesssim \num{1.5e-3} \lambda$.

These results hold when considering each effect alone. Since the deformations at different length scales are statistically independent, the systematic uncertainty in the boost factor will approximately add in quadrature. Hence, when combining the above constraints, they are expected to tighten by a factor of around $\sqrt{3} \approx 1.7$.
On the other hand, in an experimental setup one would for example measure the reflectivity in order to constrain the boost factor. Such measurement can be used to realign (tune) the dielectric disks to more optimal positions~\cite{Knirck:2019eug}. Preliminary calculations show that this could approximately soften the planarity constraints by a factor of $2$.

\begin{figure}
	\centering
	\includegraphics[width=\textwidth]{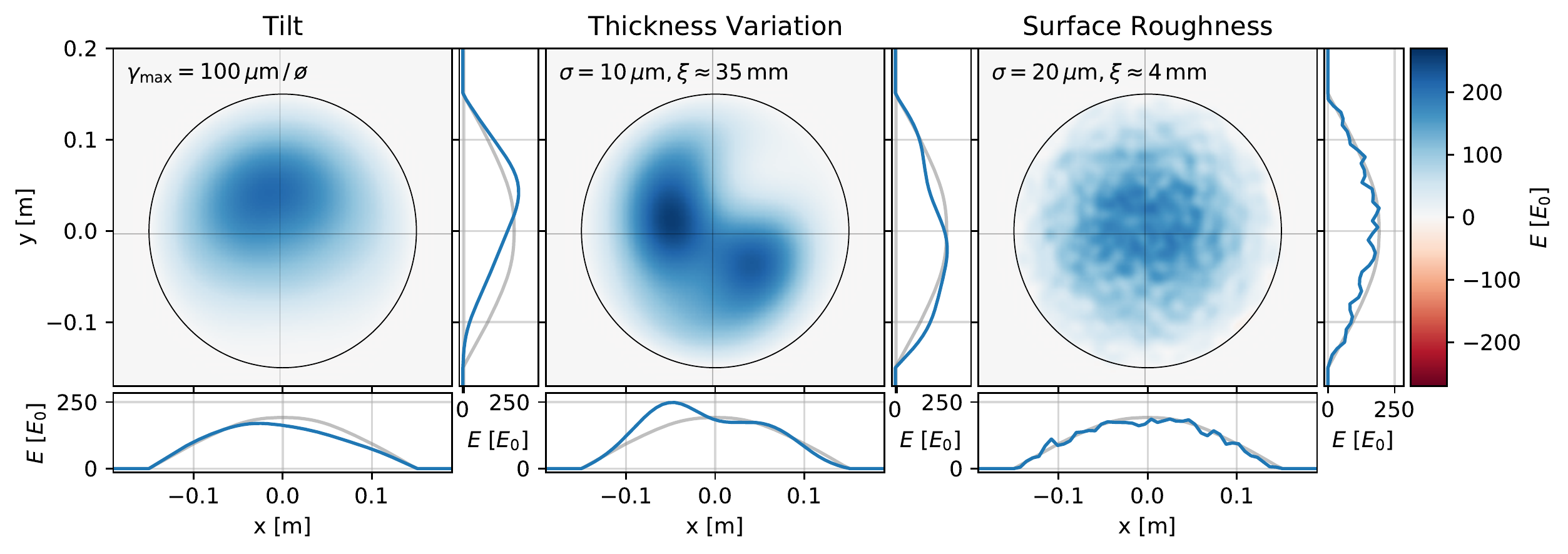}
	\vspace{-1.5em}\caption{Effect of different geometrical imperfections on the beam shape, for the 20 disk prototype booster as in figure~\ref{fig:sim:3d:result:roughness-bf} at a frequency of \SI{22.03}{\giga\hertz}. The effect of a maximum disk tilt of $\SI{100}{\micro\metre} / \o$ is shown on the left, the effect of a non-planar disks ($\sigma = \SI{10}{\micro\metre}, \xi = \SI{35}{\milli\metre}$) in the middle and the effect of a surface roughness ($\sigma = \SI{20}{\micro\metre}, \xi = \SI{4}{\milli\metre}$) on the right. The main panels show the electric field, while the sub-panels show cuts through the $x$ and $y$ axes (at $y=0,x=0$, respectively). The blue curve shows the electric field of the plotted case, while the light gray curve show the ideal case without any distortion of the booster.}
	\label{fig:sim:3d:result:roughness-bf-exampleshapes}
\end{figure}

Besides the effect on the power boost factor we show the impact on the beam shape of these effects in figure~\ref{fig:sim:3d:result:roughness-bf-exampleshapes}. It is evident that the different deformations alter the beam shapes on a similar scale as the size of distortions of the disks in the booster as expected.

\subsection{Tiled Dielectric Disks}
\label{sec:tiling}
\begin{figure}
    \centering
    \includegraphics[width=0.8\textwidth]{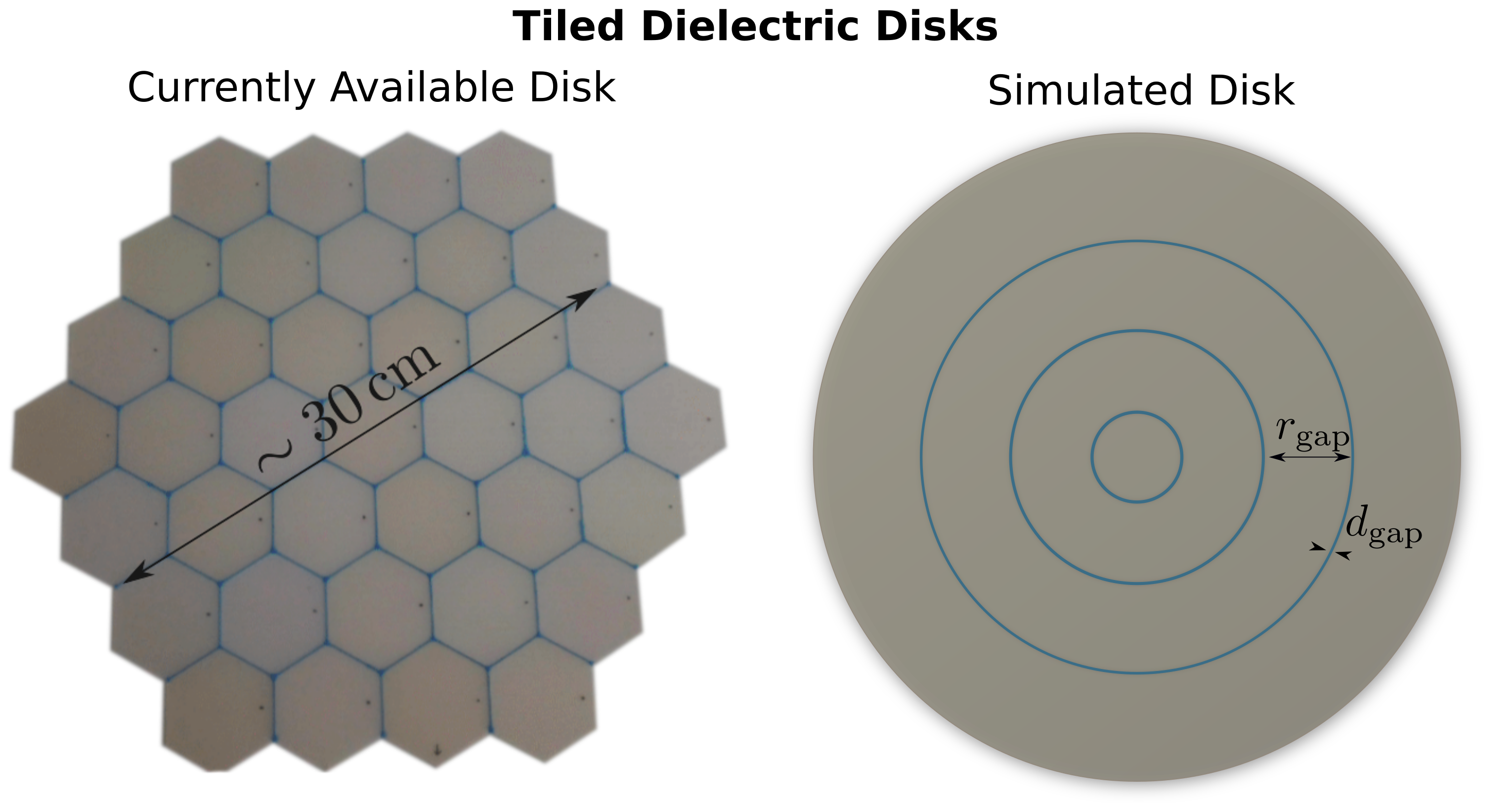}
    \caption{\textbf{Left:} Photo of the first tiled \LALO{} disk from the MADMAX collaboration~\cite{Beurthey:2020yuq}. \textbf{Right:}~Our parameterization of an azimuthally symmetric tiled disk (gluing gaps in blue). $d_{\rm gap}$ refers to the thickness of the gluing gaps, $r_{\rm gap}$ to their radial distance. The outermost tile has a \SI{1}{\centi\meter} larger width here.}
    \label{fig:tiling:parametrization}
\end{figure}

In order to achieve dielectric disks with a diameter of $\gtrsim \SI{1}{\metre}$ and low loss, the MADMAX collaboration is also investigating the possibility of gluing together smaller hexagonal patches of $\text{LaAlO}_3$ ($\epsilon \approx 24$) wavers with a diameter of around \SI{5}{\centi\metre}~\cite{Brun:2019lyf,Beurthey:2020yuq}. 
The gaps between the tiles are filled with glue ($\epsilon \approx 5$, similar to Stycast 2850FT~\cite{Halpern:86}) and have a thickness of around $\SI{0.2}{\milli\metre}$, cf.~figure~\ref{fig:tiling:parametrization}\,(left). In this section we present a first study of the impact of such a tiling on the prototype and full-scale MADMAX benchmark boost~factors.

The large difference between dielectric constants on short scales across the glued gaps invalidates the assumption of zero net charge and leads to polarization effects as already seen in~\cite{Knirck:2019eug}.
In order to apply the formalism described above, we need to derive a set of eigenmodes of the tiled disks. This can be done semi-analytically with a transfer matrix formalism~\cite{yeh2008essence}, but tends to become numerically unstable due to, again, the large relative differences in dielectric constants. This is outside the scope of this work and is left for future~studies.

However, we can efficiently simulate an azimuthally symmetric geometry with the 2D3D~FEM approach introduced in~\cite{Knirck:2019eug}.
Therefore, here we consider azimuthally symmetric, concentric tiles as shown in figure~\ref{fig:tiling:parametrization}\,(right). The parameter $r_{\text{gap}}$ describes the radial distance between two tiles and the gap thickness between two tiles is set to $d_{\text{gap}}=\SI{0.2}{\milli\metre}$.
For the prototype we set $r_{\rm gap} = \SI{4}{\centi\metre}$ to approximate the structure shown in figure~\ref{fig:tiling:parametrization}\,(left) (corresponding to three gluing gaps for $\o  = \SI{30}{\centi\metre}$, the outermost tile has a width of $\approx \SI{5}{\centi\metre}$). For the full-scale MADMAX setup we assume disks with $r_{\rm gap} = \SI{6}{\centi\metre}$, corresponding roughly to the largest possible diameter of \LALO{} crystals with currently available crystal growing techniques~\cite{LALOsize:website} (eight gluing gaps for $\o = \SI{1}{\metre}$, width of the outermost tile $\approx \SI{5}{\centi\metre}$).

\begin{figure}
    \centering
    \includegraphics[width=\textwidth]{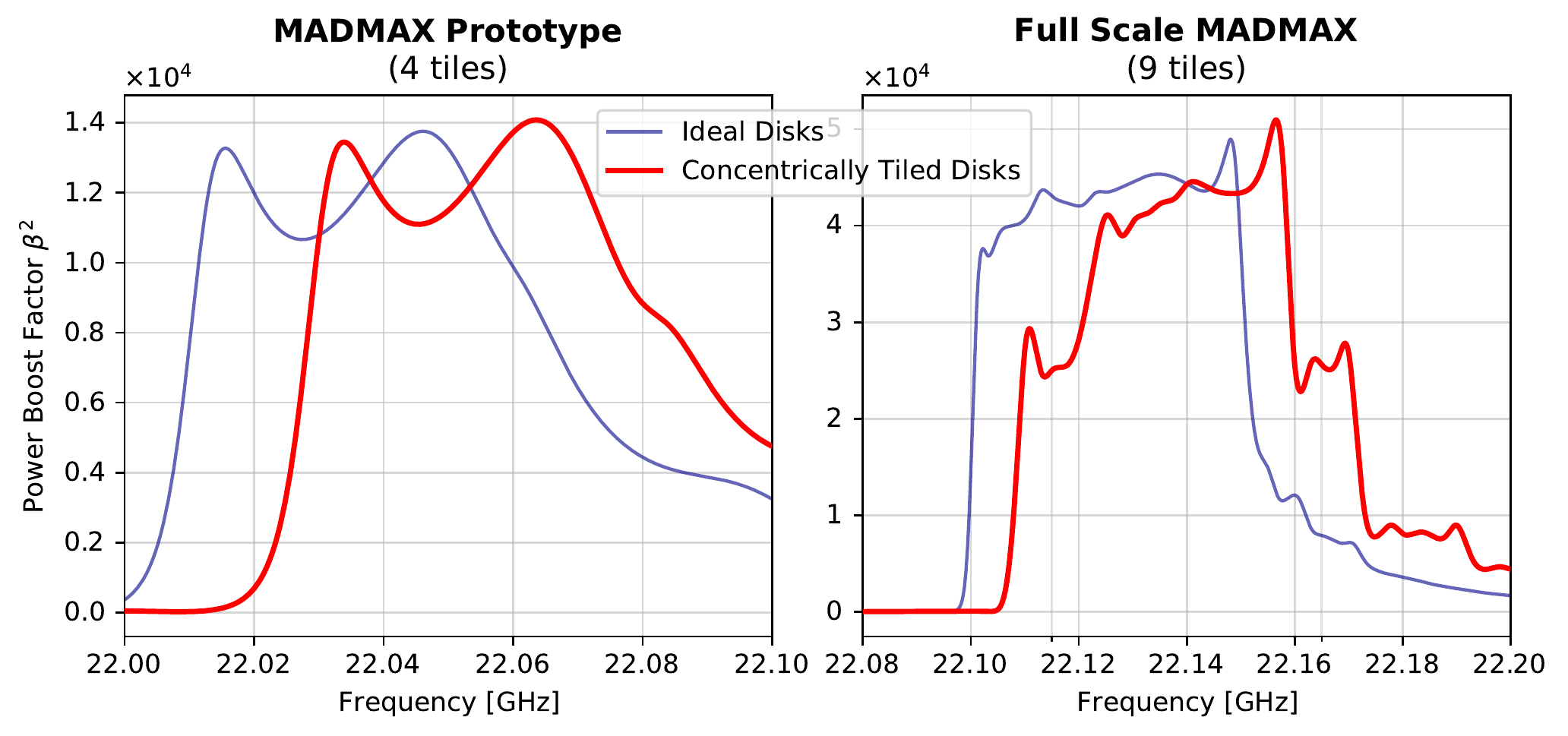}
    \caption{Effect of concentric disk tiling on the benchmark boost factors with a bandwidth of around \SI{50}{\mega\hertz} at a frequency of around \SI{22}{\giga\hertz}. We show the power boost factors $\beta^2$ without tiling (matching figure~\ref{fig:ideal}) in blue, the ones for the considered concentric tiling in red. 
    The gaps between the tiles are $d_{\rm gap} = \SI{0.2}{\milli\metre}$ thick and filled with a $\epsilon=5$ glue. The curves refer to the total emitted power.
    \textbf{Left:}~Power boost factor for the case of 20 disks with a diameter of \SI{30}{\centi\meter} as in the proposed MADMAX prototype (antenna $w_0 = \SI{10}{\centi\meter}$), with a radial distance of $r_{\rm gap} = \SI{4}{\centi\metre}$ between 4 individual tiles. \textbf{Right:}~Power boost factor for the case of 80 disks with a diameter of \SI{1}{\meter} as in the proposed MADMAX final experimental phase (antenna $w_0 = \SI{30}{\centi\meter}$), with a distance of $r_{\rm gap} = \SI{6}{\centi\metre}$ between 9  individual tiles.
    }
    \label{fig:tiling:boost_factors}
\end{figure}

Figure~\ref{fig:tiling:boost_factors} shows the result of this calculation for the prototype~(left) and full-scale~(right) MADMAX benchmark boost factors analogously to figure~\ref{fig:ideal}.
First, we see that the achievable power boost is only mildly reduced at the level of a few percent compared to the ideal 3D calculation in terms of total emitted power.
In addition, the boost factors of the untiled and tiled systems are shifted against each other in frequency. This is consistent with the expectation of additional transverse momentum to the electromagnetic wave obtained from the tiling structure. The shift is much smaller than in the case where each tile would be totally electromagnetically decoupled from each other. In this case the shift according to eq.~\eqref{eq:sim:freqshift} would naively increase by a factor of $\left(\o / r_{\rm gap}\right)^2 \approx 60$ (for the prototype) and $\approx 300$ (for the full-scale experiment).

\begin{figure}
    \centering
    \includegraphics[width=\textwidth]{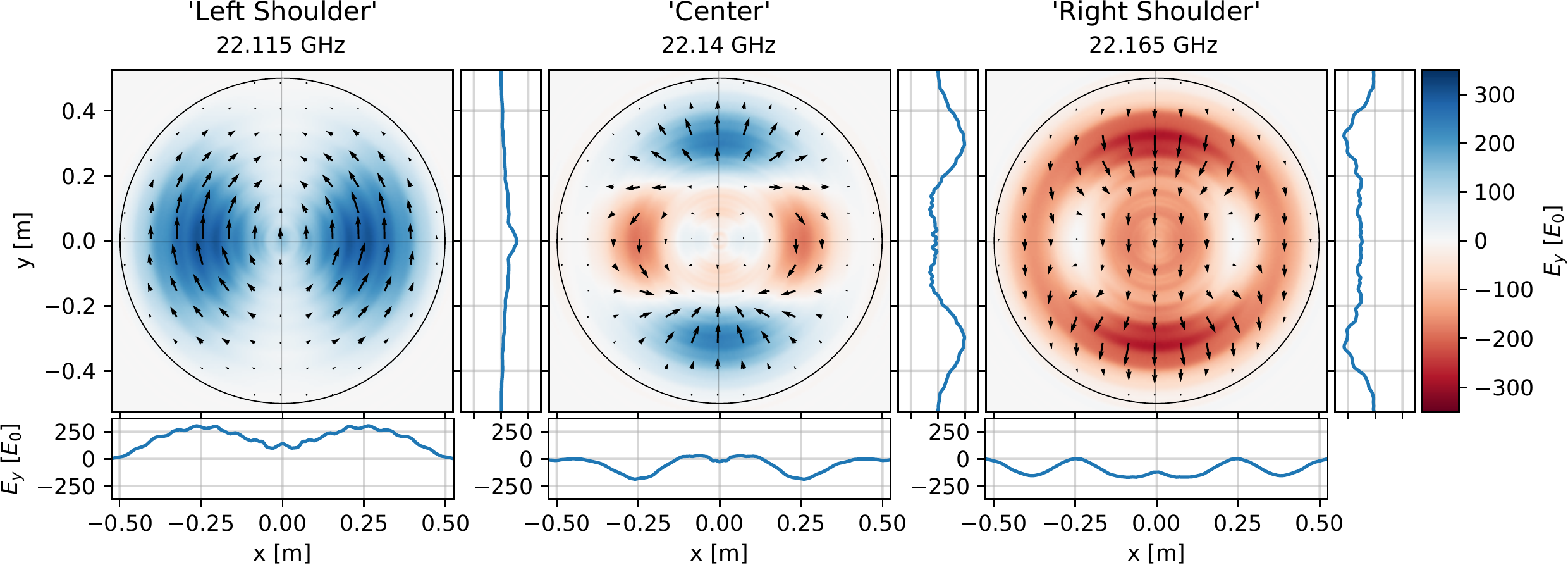}
    \caption{Beam shapes of emitted fields for azimuthally symmetric tiled disks and the full-scale MADMAX booster with 9 concentric tiles, at frequencies of \SI{22.115}{\giga\hertz} (`left shoulder' in figure~\ref{fig:tiling:boost_factors}\,(right)), \SI{22.14}{\giga\hertz} (`center' in figure~\ref{fig:tiling:boost_factors}\,(right)) and \SI{22.165}{\giga\hertz} (`right shoulder' in figure~\ref{fig:tiling:boost_factors}\,(right)). We show the physical fields at a fixed instant of time. Color denotes strength of the $y$-component of the electric field, arrows indicate strength and direction of the field. The frequencies considered here are denoted by additional frequency ticks on the horizontal axis in figure~\ref{fig:tiling:boost_factors}\,(right).}
    \label{fig:tiling:beamshapes}
\end{figure}

Next, we consider the emitted beam shape. For the MADMAX prototype the emitted power can still be received with a high efficiency of $> 90\,\%$ using the Gaussian beam antenna discussed above.
However, for the full scale MADMAX we find that the beam shape is significantly altered due to polarization effects caused by the tiling.
This is demonstrated in figure~\ref{fig:tiling:beamshapes} where we show the emitted beam shapes of the final scale MADMAX booster at representative frequencies for the full-scale MADMAX boost factor as in figure~\ref{fig:tiling:boost_factors}~(right). The electric fields have a non-negligible $x$-component, although the external magnetic field is polarized in $y$-direction, $\bm{B}_{\rm e} \propto \hat{\bm{e}}_y$.
At the lower frequency the beam shape is approximately proportional to $\cos \phi \, \hat{\bm {e}}_\phi$, at the higher frequency it is dominated by a $\sin \phi \, \hat{\bm {e}}_r$ component. At intermediate frequencies it contains both polarizations but at arbitrary phase.
Adding them in phase would give a field polarized in $y$-direction as desired, i.e., contributions from both $r$ and $\phi$ polarizations appear shifted with respect to each other in frequency.
The $x$-component always obeys a quadruple structure as we have already seen in~\cite{Knirck:2019eug}, cf.~also appendix~\ref{app:tilingrels}.
Fields $\propto \hat{\bm {e}}_r$ are orthogonal and $\propto \hat{\bm {e}}_\phi$ parallel to the glue gaps and hence need to obey different electromagnetic boundary conditions.
Our observations are consistent with $r$ and $\phi$ polarized waves therefore having different propagation constants within the booster analogous to propagation in e.g.~anisotropic media.

Using the same Gaussian antenna as proposed in the previous sections would therefore reduce the antenna coupling by roughly a factor of $2$ or more depending on frequency for the considered full scale MADMAX setup. However, this reduction may be mitigated to some extent 
by optimization of the antenna shape or disk tiling geometry, as polarization effects may be reduced by using the proper orientation of gaps and shape of tiles.
Since resolving the modes for a tiled disk is numerically challenging as described above, studies on alternative tiling designs are not presented here and left for future work.

In addition, reducing the gap size can reduce tiling effects.
Also, tiling effects are reduced when decreasing the difference of dielectric constants between the glue and the disks by, for example, using a higher-$\epsilon$ glue. Alternatively, it is possible to resort to dielectrics with lower dielectric constant $\epsilon$, which can also be grown to larger diameters.
For example, using sapphire ($\epsilon \approx 9$) instead of lanthanum aluminate ($\epsilon \approx 24$) could thus be a realistic alternative to circumvent significant tiling effects, while reducing the power boost factor by acceptable $\approx 30\,\%$. 
We also note that other communities have experience in building meter-scale telescope lenses with high accuracy and without the need of tiling~\cite{Barto:2017}. 
MADMAX could potentially compensate a reduction in $\beta^2$ to some extent by using proportionally more dielectric disks, corresponding to a respective increase in axion-photon conversion volume~\cite{millar2017dielectric}. Finally, it is noted that MADMAX sensitivity estimates, e.g.~in~\cite{Brun:2019lyf}, use a conservative system noise temperature of \SI{8}{\kelvin}, which can likely be improved for example by using traveling wave parametric amplifiers~\cite{Planat_2020,Beurthey:2020yuq} and would allow for using smaller boost factors.

The above studies only present first estimates for a specific case of tiling. More detailed studies are underway to understand the modes of a tiled booster and the dependencies on the tiling design (such as glue thickness, orientation of tiling gaps, etc.) but also on frequency, on boost factor bandwidth, on the disk diameter and on other parameters. They will provide a clearer picture of the optimal disk design for the full-scale MADMAX booster.

\section{Summary and Conclusion}\label{sec:conclusion}

In this paper we have studied 3D effects in dielectric haloscopes 
in terms of independently propagating booster eigenmodes.
We have derived expected beam shapes for the MADMAX dielectric haloscopes for the first time.
The electromagnetic fields inside the booster are not well described by a plane wave, as in previous 1D calculations.
However, for finite sized, isotropic and perfectly flat disks the dominant contribution is from the fundamental mode that has a coupling efficiency of $69\,\%$ to the axion-induced electric field. This mode can be well received using Gaussian beam quasi-optics~\cite{goldsmith1998quasioptical} matched to a Gaussian beam with a beam waist radius of $w_0 \approx \o/3$ focused at the front-most disk of the booster. This can be achieved by using a Gaussian beam horn antenna and one or more focusing mirrors, see~e.g.~\cite{goldsmith1998quasioptical,Brun:2019lyf,Beurthey:2020yuq}.

\begin{table}[b]
    \centering
	\makebox[\textwidth][c]{\begin{tabular}{lcc} 
		\toprule
		 & {MADMAX Prototype}& {Full-scale MADMAX} \\
		 & {20 disks, $\o = \SI{30}{\centi\metre}$}& {80 disks, $\o = \SI{1}{\metre}$} \\
		\midrule 
		Antenna Beam Shape * & \multicolumn{2}{c}{Gaussian $w_0 \approx \o/3$}  \\
		\midrule
		Transverse Axion Velocity & 
		$v_a < \num{1.5e-2} c$ & $v_a < \num{5e-3} c$ \\
		Transverse $B$-Field Homogeneity & \multicolumn{2}{c}{$h \lesssim 10\,\%$} \\[0.5ex]
		Disk Tilts & \multicolumn{2}{c}{$\gamma \lesssim \SI{100}{\micro\metre} / \o$} \\
		Disk Planarity ($\xi \approx \o/10$) & 
		\multicolumn{2}{c}{$\lesssim \SI{20}{\micro\metre}$ (min-to-max)}\\
		Disk Surface Roughness ($\xi \approx \lambda/4$) & \multicolumn{2}{c}{$\lesssim \SI{100}{\micro\metre}$ (min-to-max)}
		\\[0.2ex]
		\begin{tabular}{@{}l@{}}Concentric Tiling \\ ${(d_{\rm gap} = \SI{0.2}{\milli\metre}, \epsilon_{\rm gap} = 5)}$\end{tabular} 
		& (4 radial tiles: ok) & \begin{tabular}{@{}c@{}} (9 radial tiles: \\ coupling reduced) \end{tabular}  \\
		\bottomrule
		\multicolumn{3}{r}{\footnotesize * for coupling to the fundamental mode}
	\end{tabular}}
    \caption{Summary of requirements for MADMAX dielectric haloscopes derived in this paper such as to leave the benchmark boost factor ($\SI{50}{\mega\hertz}$ bandwidth at $\SI{22}{\giga\hertz}$ corresponding to $m_a \approx \SI{90}{\micro\electronvolt}$) unchanged at the level of $20\,\%$ or below.}
    \label{tab:constraints}
\end{table}
Moreover, we have derived analytical expressions to quantify the impact on dielectric haloscopes from non-ideal effects such as non-zero axion velocity, magnetic field inhomogeneity and geometrical disk inaccuracies. We have also deduced explicit requirements for the MADMAX dielectric haloscopes for a benchmark boost factor at \SI{22}{\giga\hertz} ($m_a \approx \SI{90}{\micro\electronvolt}$) and bandwidth of \SI{50}{\mega\hertz}.
Table~\ref{tab:constraints} summarizes these parameters for both the MADMAX prototype and the full-scale MADMAX booster. All values corresponding to the non-ideal booster reduce the boost factor by less than $20\,\%$ compared to the ideal 3D case. 
Realistic values for axion velocities and magnetic field inhomogeneities are mostly unproblematic for MADMAX. However, geometrical inaccuracies of the dielectric disks, such as tilts, non-planarities and surface roughness, cause phase errors, mode mixing and effective losses, and therefore lead to important design constraints.
For fixed relative boost factor bandwidth corresponding to \SI{50}{\mega\hertz} at \SI{22}{\giga\hertz}, the results hold approximately independent of the disk number in the considered range between $20$ and $80$ disks, scale with disk diameter~$\o$ as indicated and can be generalized to other frequencies, i.e., axion masses, by appropriate scaling with the wavelength~$\lambda$.
We also have shown that concentric tiling does not reduce the boost factor significantly but shifts it to higher frequencies and can affect the beam shape.
Future studies will incorporate polarization effects caused by anisotropic dielectric constants and more realistic tiling designs, such as hexagonally tiled disks.

These results are of significant importance for the experimental design of the MADMAX booster and provide first quantitative design goals for booster manufacturing. In addition, the results may be applicable to other dielectric haloscopes and similar setups such as {ORPHEUS}~\cite{Rybka:2014cya,cervantes2019orpheus}, DALI~\cite{DeMiguel-Hernandez:2020mon} and LAMPOST~\cite{Baryakhtar:2018doz}.

\appendix

\acknowledgments
This publication is based on results from the doctoral theses of SK~\cite{Knirck:2020} and JS~\cite{SchtteEngel:450146}.
The authors thank Alexander Millar for inspiring discussions and comments.
MADMAX is supported by the Deutsche Forschungsgemeinschaft (DFG, German Research Foundation) under Germany's Excellence Strategy -- EXC 2121 ``Quantum Universe'' -- 390833306. 
This work was funded by DFG via the Collaborative Research Center ``Neutrinos und Dunkle Materie in der Astro- und Teilchenphysik'', SFB 1258.
This research was supported by the Munich Institute for Astro- and Particle Physics (MIAPP) which is funded by DFG under Germany's Excellence Strategy -- EXC-2094 -- 390783311.
AS is supported by DFG under project~number~441532750.

\section{Comparison of Numerical Methods}

\subsection{Ideal 3D Booster}

\label{app:comparision}

\begin{figure}[b]
	\centering
	\includegraphics[width=0.9\textwidth]{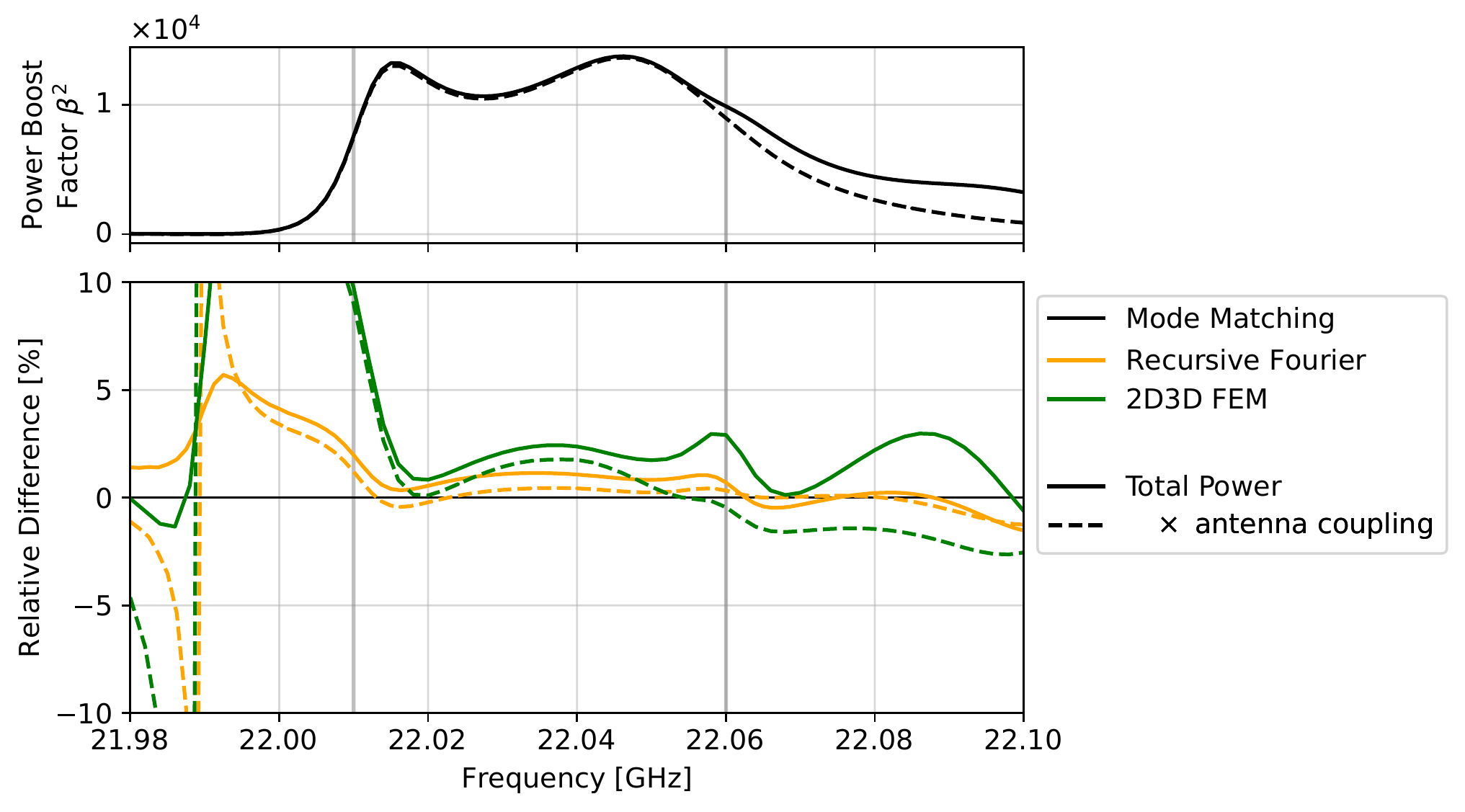}
	\caption{Comparison of (solid curves) and boost factor times antenna coupling (dashed curves) calculated with the three different numerical methods for the MADMAX prototype benchmark boost factor as in figure~\ref{fig:ideal}\,(left). The upper panel shows the boost factor calculated with the Mode Matching method for orientation. The lower panel shows the relative difference between the result from Mode Matching and either the result from Recursive Fourier Propagation (yellow) or 2D3D~FEM (green).}
	\label{fig:sim:3d:ideal:powerboostcomp}
\end{figure}

In order to verify our numerical methods against each other, we have compared the result from Mode Matching with the corresponding results from the Recursive Fourier Propagation and 2D3D~FEM methods introduced in~\cite{Knirck:2019eug}.
Figure~\ref{fig:sim:3d:ideal:powerboostcomp} shows this comparison for the MADMAX prototype benchmark boost factor discussed in this paper.
The 2D3D~FEM method solves the full vectorized wave eq.~\eqref{eq:sim:3d:waveequation-td}, while the other methods assume a scalar diffraction theory and neglect free charges in the following by setting $\nabla \cdot \bm{E} = 0$ here. The Mode Matching method in addition neglects higher modes, here $m > 5, \ell > 2$.
The lower panel shows the relative difference between the results from Recursive Fourier Propagation and 2D3D~FEM against the result from the Mode Matching method, while in the upper panel the boost factor obtained with Mode Matching is shown for orientation.
The systematic differences are likely due to the simplifying assumptions of the Recursive Fourier and Mode Matching methods. Most prominently, the boost factors obtained by Recursive Fourier Propagation and FEM are typically higher than results from Mode Matching. This is expected, since the Mode Matching method neglects higher modes which may carry additional power.
The differences are largest in the regions where the boost factor itself is small.
The boost factors are consistent up to percent level within the boost factor bandwidth. This does not significantly affect sensitivity and therefore is sufficient for this study.

It should be noted that the largest deviations indeed are outside the \SI{50}{\mega\hertz} range of the boost factor, where higher modes contribute.
Analogous results have been obtained for the MADMAX prototype within its designated frequency range at $\nu = (18, 20, 22, 24)\,\si{\giga\hertz}$ and at \SI{22}{\giga\hertz} for different boost factor bandwidths of $(5, 10, 20, 50, 100, 250)\,\si{\mega\hertz}$. In addition, the comparison has also been performed for the full-scale MADMAX setup for the $\SI{50}{\mega\hertz}$ benchmark boost factor at \SI{22}{\giga\hertz} shown in figure~\ref{fig:ideal}\,(right), leading to analogous results.

The agreement shows for the ideal 3D dielectric haloscopes consisting of multiple finite-sized disks tuned to a boost factor over a bandwidth $10^{-3} \nu$ the simplifying physics assumptions of the Mode Matching and Recursive Fourier Propagation methods are valid, i.e., a relatively low number of modes (in our case $4$) is sufficient to approximate the fields inside the system and a scalar diffraction theory neglecting effects from free charges is sufficient.

\subsection{Non-Ideal Booster}
\label{app:comparision-non-ideal}
\begin{figure}
	\centering
	\includegraphics[width=0.9\textwidth]{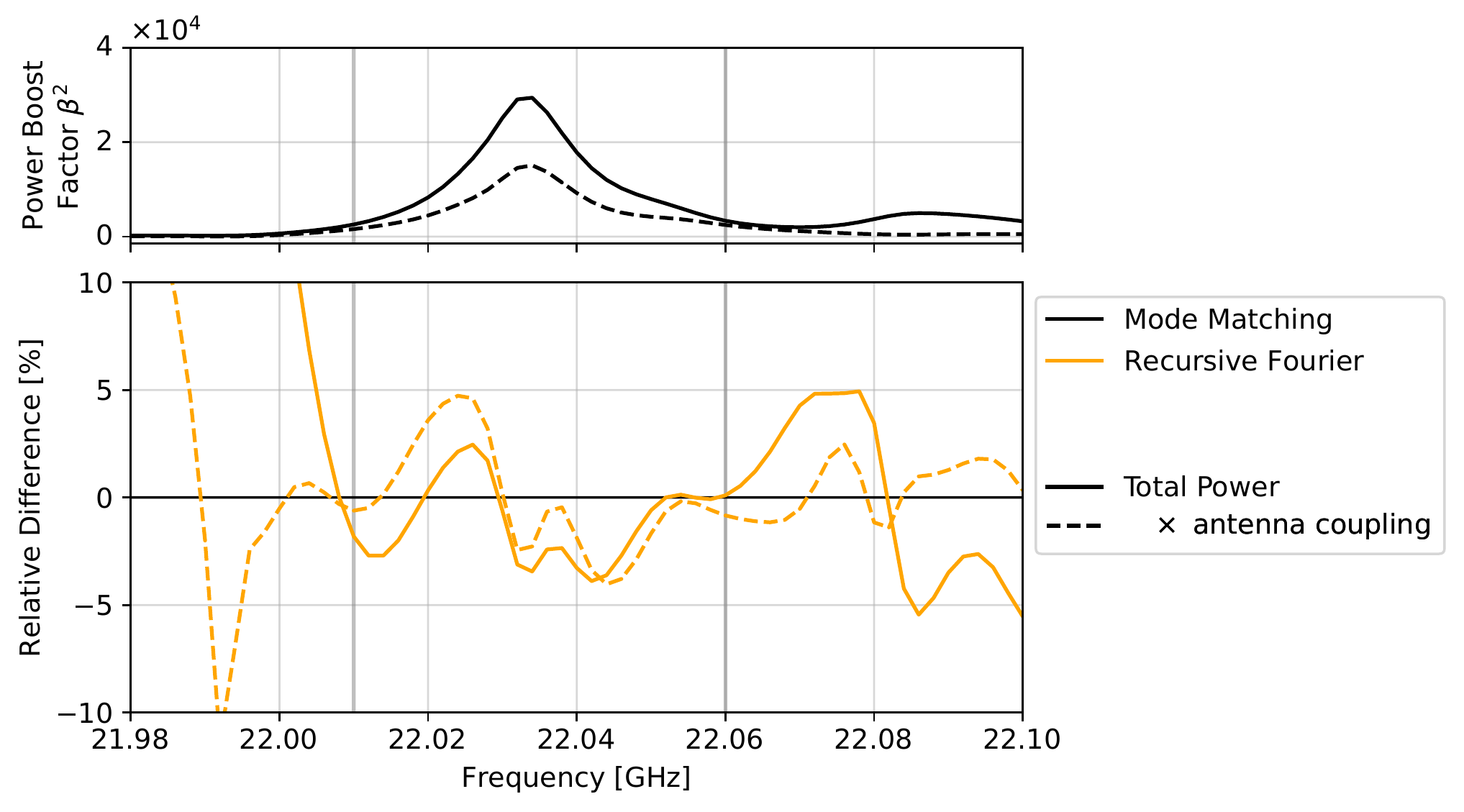}
	\caption{Comparison of boost factors (solid curves) and boost factor times antenna coupling (dashed curves) calculated with Recursive Fourier Propagating and Mode Matching, corresponding to a boost factor from the 20 disk MADMAX prototype as in figure~\ref{fig:ideal}\,(left), but now with random planarities of $\sigma = \SI{10}{\micro\metre}, \xi = \SI{35}{\milli\metre}$ for all disks. The upper panel shows the boost factor calculated with the Mode Matching method for orientation. The lower panel shows the relative difference between the result from Mode Matching and the result from Recursive Fourier Propagation.
	}
	\label{fig:sim:3d:results:roughness:method-comparision}	
\end{figure}

\begin{figure}
    \centering
    \includegraphics[width=0.95\textwidth]{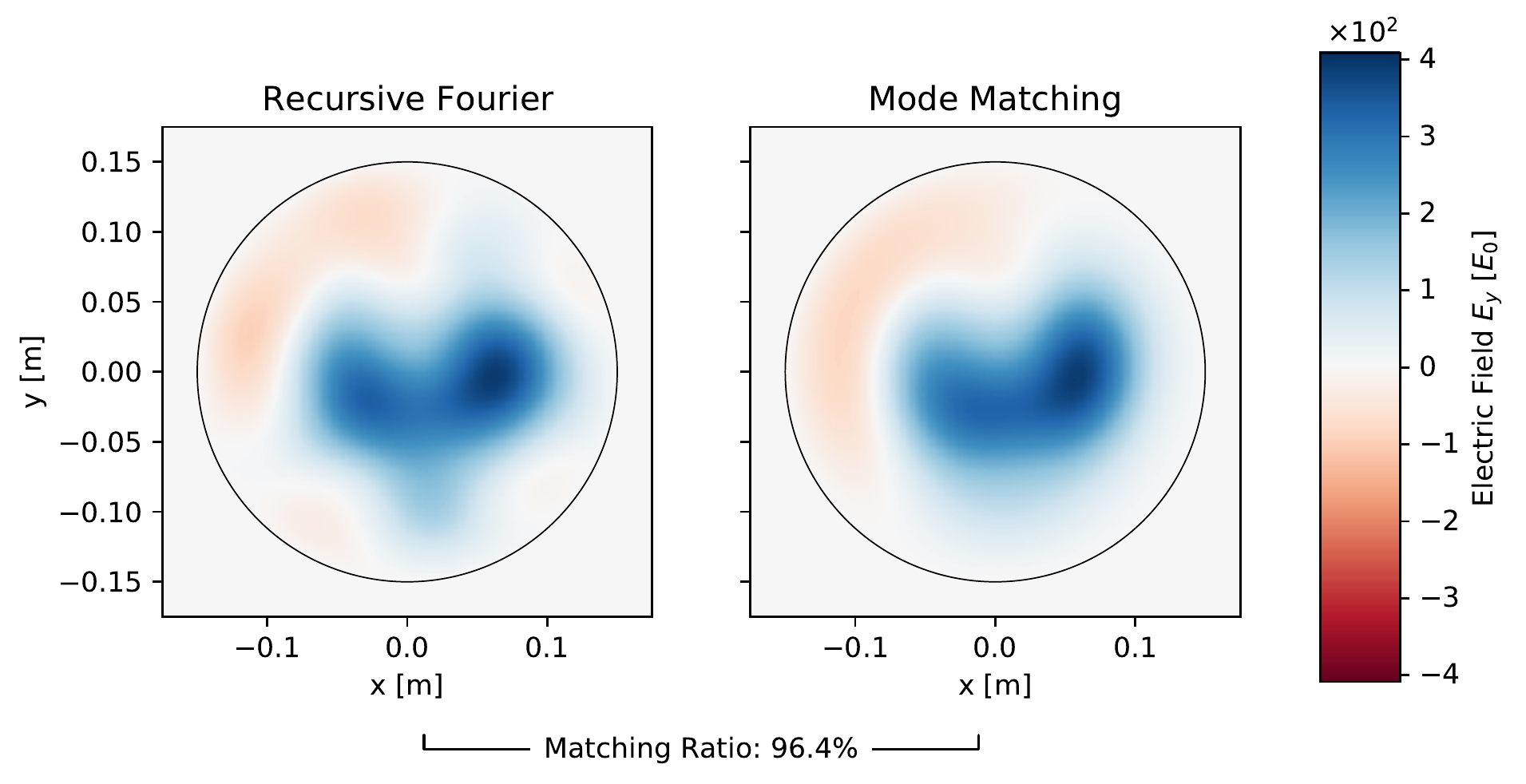}
    \caption{Comparison of beam shapes obtained with the Recursive Fourier Propagation method 
		and Mode Matching 
		for the boost factor from figure~\ref{fig:sim:3d:results:roughness:method-comparision} at the maximum boost factor.
		The matching ratio quantifies how much power an antenna would receive from one beam, if the antenna is matched perfectly to the other beam in the comparison.}
    \label{fig:sim:3d:results:roughness:method-comparision-beam-shape}
\end{figure}

The calculations presented in section~\ref{sec:nonideal_system} are not feasible with the 2D3D~FEM method, since the azimuthal symmetry is broken for these non-ideal boosters. 
Therefore, explicit numerical confirmation of the scalar diffraction theory for the non-ideal booster remains for future work. However, the scalar theory holds in the limit where $\mathbf{k} \perp \mathbf{E}$, which is still a good approximation for the MADMAX setups discussed here.
In addition, we have compared results from Mode Matching with results from Recursive Fourier Propagation for the 20 disk benchmark boost factor as in figure~\ref{fig:ideal}\,(left). 
Figure~\ref{fig:sim:3d:results:roughness:method-comparision} shows the same result for an exaggerated pessimistic planarity of $\sigma = \SI{10}{\micro\metre}$ at a scale of $\xi = \SI{35}{\milli\metre}$.
The percent level differences are irrelevant for sensitivity estimates.
We also show a comparison for the planarity calculation between the beam shapes obtained at the frequency with maximum boost in figure~\ref{fig:sim:3d:results:roughness:method-comparision-beam-shape}.
The observed differences are at smaller scales than the considered modes, i.e., mainly arise due to the fact that the Mode Matching method is neglecting higher modes. The differences can be reduced when taking into account more~modes.
Analogous results are obtained for velocity effects, magnetic field inhomogeneities and tilts.

\section{Analytical Coupling Efficiencies for Velocity Effects}
\label{app:velocity}
For $\ell = 0$ the coupling efficiencies between the axion-induced field and the modes described in section~\ref{sec:modes} is found to change with transverse velocity $v_{a,\parallel}$ as
\begin{equation}
    \frac{\Delta \eta_{m0}}{\eta_{m0}} = 1 -
    \frac{j_{0,m} \left[m_a v_{a,\parallel} R J_0\left(j_{0,m}\right) J_1(m_a v_{a,\parallel} R)-j_{0,m} J_1\left(j_{0,m}\right) J_0(m_a v_{a,\parallel} R)\right]}{J_1\left(j_{0,m}\right) \left( \left[ m_a v_{a,\parallel} R\right]^2-\left(j_{0,m}\right){}^2\right)}.
\end{equation}

\section{Explicit Mode Mixing Matrix Calculations for Transverse Disk Thickness Variations}
\label{app:mxing_thicknessvar}

\begin{figure}
	\centering
	\includegraphics[width=0.95\textwidth]{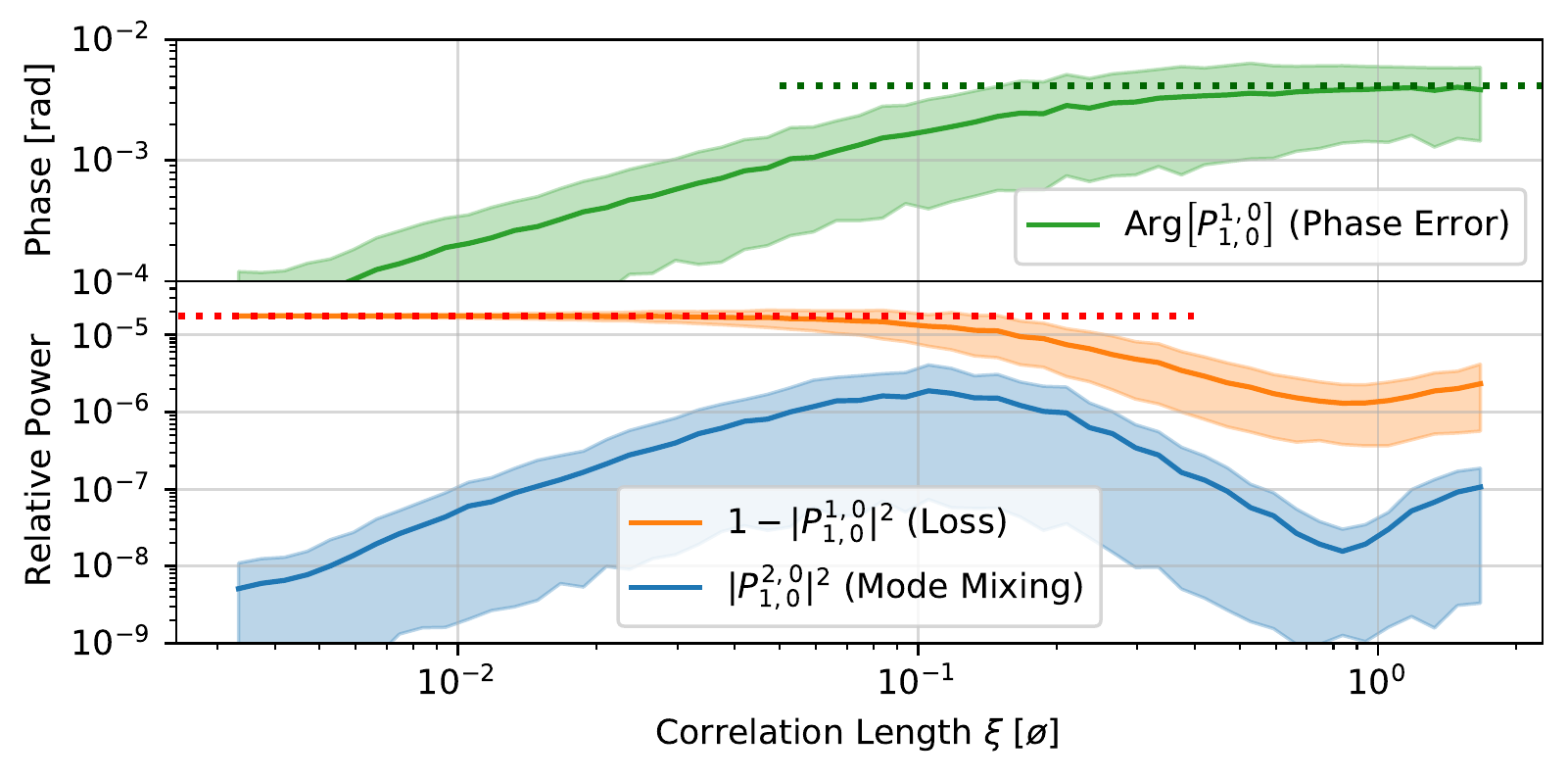}
	\caption{Behavior of various properties of the mixing matrix $P$ as a function of the correlation length $\xi$ of \SI{10}{\micro\metre} thickness variations in a vacuum gap. The top panel shows the phase shift of the fundamental mode due to the thickness variation ${\rm Arg} \left[P_{1,0}^{1,0} \right]$ in green. The green horizontal dotted line corresponds the expectation from eq.~\eqref{eq:sim:3d:results:tilt:mixing-phase-out}.
	The lower panel shows the lost power in $P_{1,0}^{1,0}$, in red and the magnitude of the off-diagonal coupling $P_{1,0}^{2,0}$ in blue. The horizontal dotted line corresponds to the expectation from eq.~\eqref{eq:sim:3d:results:tilt:mixing-loss}.
	The solid lines correspond to the expectation value after averaging over many realizations of thickness variations. The shaded regions correspond to the range of \SI{70}{\percent} of the realizations of such thickness variations.
	}
	\label{fig:sim:3d:result:roughness:errors-scaling}
\end{figure}

We have verified the initial estimates in section~\ref{sec:nonideal_system:thicknessvar} by calculating the mixing matrix $P$ numerically for many realization of thickness variations $\Delta z(r,\phi)$ and observing how it changes with $\xi$.
Figure~\ref{fig:sim:3d:result:roughness:errors-scaling} shows the result of such a calculation for $\sigma = \SI{10}{\micro\metre}$. Other $\sigma$ give analogous results.
Here we used $\approx 10^3$ samples for the thickness variation at each scale. 
Each sample corresponds to one realization of thickness variation.
Higher number of samples make the lines in the figure smoother.
Solid lines correspond to the average results while the shaded regions around them to the region in which $70\,\%$ of all results ended up.
The top panel shows how the phase error for the fundamental mode, $\Phi_m = {\rm Arg} \left[P_{m,0}^{m,0}\right]$, changes with $\xi$. 
The phase increases with $\xi$ and approaches the value expected for a misplacement of $\SI{10}{\micro\metre}$ (horizontal dotted line) when $\xi$ approaches the disk diameter $\xi \approx \o$.
The orange line in the lower panel shows the lost power in the fundamental mode, $1 - \left|P_{1,0}^{1,0}\right|^2$,  as a function of $\xi$. 
Losses are large for low $\xi$ and approach the value expected from eq.~\eqref{eq:sim:3d:results:tilt:mixing-loss} (horizontal dotted line).
Lastly, the blue curve in the bottom panel shows the magnitude of the mixing between the fundamental and the second mode, $\left|P_{1,0}^{2,0}\right|^2$, as a function of $\xi$. It is maximized near the region where $\xi$ is comparable to the length scale of the modes.
Since the length scales for the $m = 1\,...\,4$ modes are similar, analogous results are obtained for the other relevant modes of the dielectric haloscope.

\section{Allowed Fields for Azimuthal Symmetry and Linearly Polarized Source Term}
\label{app:tilingrels}
In the case of an azimuthally symmetric geometry and a linearly polarized external $B$-field in the $y$-direction, we can use the 2D3D approach introduced in~\cite{Knirck:2019eug}. 
The total solution is obtained as a superposition
\begin{eqnarray}
\bm{E}=\tilde{\bm{E}}^{+}e^{i\phi}+\tilde{\bm{E}}^{-}e^{-i\phi},
\label{eq:3DSolutions_Edecomposition}
\end{eqnarray}
where the fields $\tilde{\bm{E}}^{+}$ and $\tilde{\bm{E}}^{-}$ are calculated numerically as described in~\cite{Knirck:2019eug}.
It is important to notice that the $\pm$ solutions obey the relations
\begin{eqnarray}
\tilde{E}_{r}^-&\approx&-\tilde{E}_{r}^+,\label{eq:2D3DrelEr_1}\\
\tilde{E}_{\phi}^-&\approx&\tilde{E}_{\phi}^+,\label{eq:2D3DrelEp_1}\\
\tilde{E}_{z}^-&\approx&-\tilde{E}_{z}^+.\label{eq:2D3DrelEz_1}
\end{eqnarray}
which have not been used in~\cite{Knirck:2019eug}.

Using eq.~\eqref{eq:3DSolutions_Edecomposition}--\eqref{eq:2D3DrelEz_1} the $\hat{\bm{e}}_r$, $\hat{\bm{e}}_\phi$ and $\hat{\bm{e}}_z$ contributions to the total $\bm{E}$ are
\begin{eqnarray}
\bm{E}_r&\sim&\tilde{E}^+_r(r,z)\sin\phi\begin{pmatrix}
\cos\phi\\\sin\phi\\0
\end{pmatrix},\label{eq:Tiling_80Disk_interpretation_fields1}\\
\bm{E}_\phi&\sim&\tilde{E}^+_\phi(r,z)\cos\phi\begin{pmatrix}
-\sin\phi\\\cos\phi\\0
\end{pmatrix},\label{eq:Tiling_80Disk_interpretation_fields2}\\
\bm{E}_z&\sim& \tilde{E}^+_z(r,z)\sin\phi~ \hat{\bm{e}}_z.
\label{eq:Tiling_80Disk_interpretation_fields3}
\end{eqnarray}
From eq.~\eqref{eq:Tiling_80Disk_interpretation_fields1}--\eqref{eq:Tiling_80Disk_interpretation_fields3} we see that the $r$-polarized electric fields always have $\sin \phi$ and the $\phi$-polarized ones $\cos \phi$ dependence. We also note that the $x$-component of the fields obeys always a quadrupole structure. The $y$-component of the $r$ ($\phi$) component obeys a vertical (horizontal) dipole structure.

\bibliographystyle{JHEP}
\bibliography{bibliography}

\providecommand{\noopsort}[1]{}\providecommand{\singleletter}[1]{#1}%

\providecommand{\href}[2]{#2}\begingroup\raggedright\begin{thebibliography}{10}

\bibitem{PhysRevLett.38.1440}
R.~D. Peccei and H.~R. Quinn, \emph{$\mathit{CP}$ conservation in the presence
  of pseudoparticles},
  \href{https://doi.org/10.1103/PhysRevLett.38.1440}{\emph{Phys. Rev. Lett.}
  {\bfseries 38} (Jun, 1977) 1440--1443}.

\bibitem{PhysRevLett.40.223}
S.~Weinberg, \emph{A new light boson?},
  \href{https://doi.org/10.1103/PhysRevLett.40.223}{\emph{Phys. Rev. Lett.}
  {\bfseries 40} (Jan, 1978) 223--226}.

\bibitem{PhysRevLett.40.279}
F.~Wilczek, \emph{Problem of strong $\mathit{P}$ and $\mathit{T}$ invariance in
  the presence of instantons},
  \href{https://doi.org/10.1103/PhysRevLett.40.279}{\emph{Phys. Rev. Lett.}
  {\bfseries 40} (Jan, 1978) 279--282}.

\bibitem{DINE1983137}
M.~Dine and W.~Fischler, \emph{The not-so-harmless axion},
  \href{https://doi.org/https://doi.org/10.1016/0370-2693(83)90639-1}{\emph{Physics
  Letters B} {\bfseries 120} (1983) 137 -- 141}.

\bibitem{PRESKILL1983127}
J.~Preskill, M.~B. Wise and F.~Wilczek, \emph{Cosmology of the invisible
  axion},
  \href{https://doi.org/https://doi.org/10.1016/0370-2693(83)90637-8}{\emph{Physics
  Letters B} {\bfseries 120} (1983) 127 -- 132}.

\bibitem{ABBOTT1983133}
L.~Abbott and P.~Sikivie, \emph{A cosmological bound on the invisible axion},
  \href{https://doi.org/https://doi.org/10.1016/0370-2693(83)90638-X}{\emph{Physics
  Letters B} {\bfseries 120} (1983) 133 -- 136}.

\bibitem{DAVIS1986225}
R.~Davis, \emph{Cosmic axions from cosmic strings},
  \href{https://doi.org/https://doi.org/10.1016/0370-2693(86)90300-X}{\emph{Physics
  Letters B} {\bfseries 180} (1986) 225 -- 230}.

\bibitem{LYTH1992189}
D.~H. Lyth and E.~D. Stewart, \emph{Constraining the inflationary energy scale
  from axion cosmology},
  \href{https://doi.org/https://doi.org/10.1016/0370-2693(92)90006-P}{\emph{Physics
  Letters B} {\bfseries 283} (1992) 189 -- 193}.

\bibitem{Kawasaki:2014sqa}
M.~Kawasaki, K.~Saikawa and T.~Sekiguchi, \emph{{Axion dark matter from
  topological defects}},
  \href{https://doi.org/10.1103/PhysRevD.91.065014}{\emph{Phys. Rev.}
  {\bfseries D91} (2015) 065014},
  [\href{https://arxiv.org/abs/1412.0789}{{\ttfamily 1412.0789}}].

\bibitem{Fleury:2015aca}
L.~Fleury and G.~D. Moore, \emph{{Axion dark matter: strings and their cores}},
  \href{https://doi.org/10.1088/1475-7516/2016/01/004}{\emph{JCAP} {\bfseries
  1601} (2016) 004}, [\href{https://arxiv.org/abs/1509.00026}{{\ttfamily
  1509.00026}}].

\bibitem{Ringwald:2015dsf}
A.~Ringwald and K.~Saikawa, \emph{{Axion dark matter in the post-inflationary
  Peccei-Quinn symmetry breaking scenario}},
  \href{https://doi.org/10.1103/PhysRevD.93.085031,
  10.1103/PhysRevD.94.049908}{\emph{Phys. Rev.} {\bfseries D93} (2016) 085031},
  [\href{https://arxiv.org/abs/1512.06436}{{\ttfamily 1512.06436}}].

\bibitem{Fleury:2016xrz}
L.~M. Fleury and G.~D. Moore, \emph{{Axion String Dynamics I: 2+1D}},
  \href{https://doi.org/10.1088/1475-7516/2016/05/005}{\emph{JCAP} {\bfseries
  1605} (2016) 005}, [\href{https://arxiv.org/abs/1602.04818}{{\ttfamily
  1602.04818}}].

\bibitem{Borsanyietal}
S.~Borsanyi, Z.~Fodor, J.~Guenther, K.-H. Kampert, S.~D. Katz, T.~Kawanai
  et~al., \emph{Calculation of the axion mass based on high-temperature lattice
  quantum chromodynamics}, {\emph{Nature} {\bfseries 539} (Nov., 2016) 69--71}.

\bibitem{Ballesteros:2016xej}
G.~Ballesteros, J.~Redondo, A.~Ringwald and C.~Tamarit, \emph{{Standard
  Model—axion—seesaw—Higgs portal inflation. Five problems of particle
  physics and cosmology solved in one stroke}},
  \href{https://doi.org/10.1088/1475-7516/2017/08/001}{\emph{JCAP} {\bfseries
  1708} (2017) 001}, [\href{https://arxiv.org/abs/1610.01639}{{\ttfamily
  1610.01639}}].

\bibitem{PhysRevD.98.030001}
{\scshape Particle Data Group} collaboration, M.~T. {\it et al.}~[Particle
  Data~Group], \emph{Review of particle physics},
  \href{https://doi.org/10.1103/PhysRevD.98.030001}{\emph{Phys. Rev. D}
  {\bfseries 98} (Aug, 2018) 030001}.

\bibitem{Braine:2019fqb}
{\scshape ADMX} collaboration, T.~Braine et~al., \emph{{Extended Search for the
  Invisible Axion with the Axion Dark Matter Experiment}},
  \href{https://doi.org/10.1103/PhysRevLett.124.101303}{\emph{Phys. Rev. Lett.}
  {\bfseries 124} (2020) 101303},
  [\href{https://arxiv.org/abs/1910.08638}{{\ttfamily 1910.08638}}].

\bibitem{Backes:2020ajv}
{\scshape HAYSTAC} collaboration, K.~Backes et~al., \emph{{A quantum-enhanced
  search for dark matter axions}},
  \href{https://arxiv.org/abs/2008.01853}{{\ttfamily 2008.01853}}.

\bibitem{PhysRevD.85.105020}
T.~Hiramatsu, M.~Kawasaki, K.~Saikawa and T.~Sekiguchi, \emph{Production of
  dark matter axions from collapse of string-wall systems},
  \href{https://doi.org/10.1103/PhysRevD.85.105020}{\emph{Phys. Rev. D}
  {\bfseries 85} (May, 2012) 105020}.

\bibitem{Klaer:2017ond}
V.~B. Klaer and G.~D. Moore, \emph{{The dark-matter axion mass}},
  \href{https://doi.org/10.1088/1475-7516/2017/11/049}{\emph{JCAP} {\bfseries
  1711} (2017) 049}, [\href{https://arxiv.org/abs/1708.07521}{{\ttfamily
  1708.07521}}].

\bibitem{Kawasaki:2018bzv}
M.~Kawasaki, T.~Sekiguchi, M.~Yamaguchi and J.~Yokoyama, \emph{{Long-term
  dynamics of cosmological axion strings}},
  \href{https://doi.org/10.1093/ptep/pty098}{\emph{PTEP} {\bfseries 2018}
  (2018) 091E01}, [\href{https://arxiv.org/abs/1806.05566}{{\ttfamily
  1806.05566}}].

\bibitem{Gorghetto:2018myk}
M.~Gorghetto, E.~Hardy and G.~Villadoro, \emph{{Axions from Strings: the
  Attractive Solution}},
  \href{https://doi.org/10.1007/JHEP07(2018)151}{\emph{JHEP} {\bfseries 07}
  (2018) 151}, [\href{https://arxiv.org/abs/1806.04677}{{\ttfamily
  1806.04677}}].

\bibitem{millar2017dielectric}
A.~J. Millar, G.~G. Raffelt, J.~Redondo and F.~D. Steffen, \emph{{Dielectric
  Haloscopes to Search for Axion Dark Matter: Theoretical Foundations}},
  \href{https://doi.org/10.1088/1475-7516/2017/01/061}{\emph{JCAP} {\bfseries
  1701} (2017) 061}, [\href{https://arxiv.org/abs/1612.07057}{{\ttfamily
  1612.07057}}].

\bibitem{TheMADMAXWorkingGroup:2016hpc}
{\scshape MADMAX Working Group} collaboration, A.~Caldwell, G.~Dvali,
  B.~Majorovits, A.~Millar, G.~Raffelt, J.~Redondo et~al., \emph{{Dielectric
  Haloscopes: A New Way to Detect Axion Dark Matter}},
  \href{https://doi.org/10.1103/PhysRevLett.118.091801}{\emph{Phys. Rev. Lett.}
  {\bfseries 118} (2017) 091801},
  [\href{https://arxiv.org/abs/1611.05865}{{\ttfamily 1611.05865}}].

\bibitem{Brun:2019lyf}
{\scshape MADMAX} collaboration, P.~Brun et~al., \emph{{A new experimental
  approach to probe QCD axion dark matter in the mass range above 40 $\mu$eV}},
  \href{https://doi.org/10.1140/epjc/s10052-019-6683-x}{\emph{Eur. Phys. J. C}
  {\bfseries 79} (2019) 186},
  [\href{https://arxiv.org/abs/1901.07401}{{\ttfamily 1901.07401}}].

\bibitem{Horns:2012jf}
D.~Horns, J.~Jaeckel, A.~Lindner, A.~Lobanov, J.~Redondo and A.~Ringwald,
  \emph{{Searching for WISPy Cold Dark Matter with a Dish Antenna}},
  \href{https://doi.org/10.1088/1475-7516/2013/04/016}{\emph{JCAP} {\bfseries
  1304} (2013) 016}, [\href{https://arxiv.org/abs/1212.2970}{{\ttfamily
  1212.2970}}].

\bibitem{McAllister:2017lkb}
B.~T. McAllister, G.~Flower, E.~N. Ivanov, M.~Goryachev, J.~Bourhill and M.~E.
  Tobar, \emph{{The ORGAN Experiment: An axion haloscope above 15 GHz}},
  \href{https://doi.org/10.1016/j.dark.2017.09.010}{\emph{Phys. Dark Univ.}
  {\bfseries 18} (2017) 67--72},
  [\href{https://arxiv.org/abs/1706.00209}{{\ttfamily 1706.00209}}].

\bibitem{Jeong:2017hqs}
J.~Jeong, S.~Youn, S.~Ahn, J.~E. Kim and Y.~K. Semertzidis, \emph{{Concept of
  multiple-cell cavity for axion dark matter search}},
  \href{https://doi.org/10.1016/j.physletb.2017.12.066}{\emph{Phys. Lett.}
  {\bfseries B777} (2018) 412--419},
  [\href{https://arxiv.org/abs/1710.06969}{{\ttfamily 1710.06969}}].

\bibitem{Melcon:2018dba}
A.~A. Melcon et~al., \emph{{Axion Searches with Microwave Filters: the RADES
  project}}, \href{https://doi.org/10.1088/1475-7516/2018/05/040}{\emph{JCAP}
  {\bfseries 1805} (2018) 040},
  [\href{https://arxiv.org/abs/1803.01243}{{\ttfamily 1803.01243}}].

\bibitem{Baryakhtar:2018doz}
M.~Baryakhtar, J.~Huang and R.~Lasenby, \emph{{Axion and hidden photon dark
  matter detection with multilayer optical haloscopes}},
  \href{https://doi.org/10.1103/PhysRevD.98.035006}{\emph{Phys. Rev.}
  {\bfseries D98} (2018) 035006},
  [\href{https://arxiv.org/abs/1803.11455}{{\ttfamily 1803.11455}}].

\bibitem{Lawson:2019brd}
M.~Lawson, A.~J. Millar, M.~Pancaldi, E.~Vitagliano and F.~Wilczek,
  \emph{{Tunable axion plasma haloscopes}},
  \href{https://doi.org/10.1103/PhysRevLett.123.141802}{\emph{Phys. Rev. Lett.}
  {\bfseries 123} (2019) 141802},
  [\href{https://arxiv.org/abs/1904.11872}{{\ttfamily 1904.11872}}].

\bibitem{Suzuki:2015sza}
J.~Suzuki, T.~Horie, Y.~Inoue and M.~Minowa, \emph{{Experimental Search for
  Hidden Photon CDM in the eV mass range with a Dish Antenna}},
  \href{https://doi.org/10.1088/1475-7516/2015/09/042,
  10.1088/1475-7516/2015/9/042}{\emph{JCAP} {\bfseries 1509} (2015) 042},
  [\href{https://arxiv.org/abs/1504.00118}{{\ttfamily 1504.00118}}].

\bibitem{FUNK:2017icw}
{\scshape FUNK Experiment} collaboration, D.~Veberič et~al., \emph{{Search for
  hidden-photon dark matter with the FUNK experiment}},
  \href{https://doi.org/10.22323/1.301.0880}{\emph{PoS} {\bfseries ICRC2017}
  (2018) 880}, [\href{https://arxiv.org/abs/1711.02958}{{\ttfamily
  1711.02958}}].

\bibitem{Knirck:2018ojz}
S.~Knirck, T.~Yamazaki, Y.~Okesaku, S.~Asai, T.~Idehara and T.~Inada,
  \emph{{First results from a hidden photon dark matter search in the meV
  sector using a plane-parabolic mirror system}},
  \href{https://doi.org/10.1088/1475-7516/2018/11/031}{\emph{JCAP} {\bfseries
  1811} (2018) 031}, [\href{https://arxiv.org/abs/1806.05120}{{\ttfamily
  1806.05120}}].

\bibitem{BRASS:website}
``Brass website.''
  \url{http://wwwiexp.desy.de/groups/astroparticle/brass/brassweb.htm}
  [accessed 2020-11-12].

\bibitem{Raaijmakers:2019hqj}
C.~Vigo, L.~Gerchow, B.~Radics, M.~Raaijmakers, A.~Rubbia and P.~Crivelli,
  \emph{{New bounds from positronium decays on massless mirror dark photons}},
  \href{https://doi.org/10.1103/PhysRevLett.124.101803}{\emph{Phys. Rev. Lett.}
  {\bfseries 124} (2020) 101803},
  [\href{https://arxiv.org/abs/1905.09128}{{\ttfamily 1905.09128}}].

\bibitem{PhysRevLett.123.121601}
D.~J.~E. Marsh, K.~C. Fong, E.~W. Lentz, L.~\ifmmode~\check{S}\else
  \v{S}\fi{}mejkal and M.~N. Ali, \emph{Proposal to detect dark matter using
  axionic topological antiferromagnets},
  \href{https://doi.org/10.1103/PhysRevLett.123.121601}{\emph{Phys. Rev. Lett.}
  {\bfseries 123} (Sep, 2019) 121601}.

\bibitem{Zarei:2019sva}
M.~Zarei, S.~Shakeri, M.~Abdi, D.~J.~E. Marsh and S.~Matarrese, \emph{{Probing
  Virtual Axion-Like Particles by Precision Phase Measurements}},
  \href{https://arxiv.org/abs/1910.09973}{{\ttfamily 1910.09973}}.

\bibitem{10.1007/978-3-319-92726-8_16}
\emph{Abracadabra: A broadband/resonant search for axions},  in \emph{Microwave
  Cavities and Detectors for Axion Research} (G.~Carosi, G.~Rybka and K.~van
  Bibber, eds.), (Cham), pp.~135--142, Springer International Publishing, 2018.

\bibitem{BARBIERI2017135}
R.~Barbieri, C.~Braggio, G.~Carugno, C.~Gallo, A.~Lombardi, A.~Ortolan et~al.,
  \emph{Searching for galactic axions through magnetized media: The quax
  proposal},
  \href{https://doi.org/https://doi.org/10.1016/j.dark.2017.01.003}{\emph{Physics
  of the Dark Universe} {\bfseries 15} (2017) 135 -- 141}.

\bibitem{PhysRevX.4.021030}
D.~Budker, P.~W. Graham, M.~Ledbetter, S.~Rajendran and A.~O. Sushkov,
  \emph{Proposal for a cosmic axion spin precession experiment (casper)},
  \href{https://doi.org/10.1103/PhysRevX.4.021030}{\emph{Phys. Rev. X}
  {\bfseries 4} (May, 2014) 021030}.

\bibitem{Geraci:2017bmq}
{\scshape ARIADNE} collaboration, A.~A. Geraci et~al., \emph{{Progress on the
  ARIADNE axion experiment}},
  \href{https://doi.org/10.1007/978-3-319-92726-8_18}{\emph{Springer Proc.
  Phys.} {\bfseries 211} (2018) 151--161},
  [\href{https://arxiv.org/abs/1710.05413}{{\ttfamily 1710.05413}}].

\bibitem{Semertzidis:2019gkj}
Y.~K. Semertzidis et~al., \emph{{Axion Dark Matter Research with IBS/CAPP}},
  \href{https://arxiv.org/abs/1910.11591}{{\ttfamily 1910.11591}}.

\bibitem{Schutte-Engel:2021bqm}
J.~Sch\"utte-Engel, D.~J.~E. Marsh, A.~J. Millar, A.~Sekine, F.~Chadha-Day,
  S.~Hoof et~al., \emph{{Axion Quasiparticles for Axion Dark Matter
  Detection}},  \href{https://arxiv.org/abs/2102.05366}{{\ttfamily
  2102.05366}}.

\bibitem{Graham:2015ouw}
P.~W. Graham, I.~G. Irastorza, S.~K. Lamoreaux, A.~Lindner and K.~A. van
  Bibber, \emph{{Experimental Searches for the Axion and Axion-Like
  Particles}},
  \href{https://doi.org/10.1146/annurev-nucl-102014-022120}{\emph{Ann. Rev.
  Nucl. Part. Sci.} {\bfseries 65} (2015) 485--514},
  [\href{https://arxiv.org/abs/1602.00039}{{\ttfamily 1602.00039}}].

\bibitem{Irastorza:2018dyq}
I.~G. Irastorza and J.~Redondo, \emph{{New experimental approaches in the
  search for axion-like particles}},
  \href{https://doi.org/10.1016/j.ppnp.2018.05.003}{\emph{Prog. Part. Nucl.
  Phys.} {\bfseries 102} (2018) 89--159},
  [\href{https://arxiv.org/abs/1801.08127}{{\ttfamily 1801.08127}}].

\bibitem{Kim:1979if}
J.~E. Kim, \emph{{Weak Interaction Singlet and Strong CP Invariance}},
  \href{https://doi.org/10.1103/PhysRevLett.43.103}{\emph{Phys. Rev. Lett.}
  {\bfseries 43} (1979) 103}.

\bibitem{Shifman:1979if}
M.~A. Shifman, A.~I. Vainshtein and V.~I. Zakharov, \emph{{Can Confinement
  Ensure Natural CP Invariance of Strong Interactions?}},
  \href{https://doi.org/10.1016/0550-3213(80)90209-6}{\emph{Nucl. Phys.}
  {\bfseries B166} (1980) 493--506}.

\bibitem{Dine:1981rt}
M.~Dine, W.~Fischler and M.~Srednicki, \emph{{A Simple Solution to the Strong
  CP Problem with a Harmless Axion}},
  \href{https://doi.org/10.1016/0370-2693(81)90590-6}{\emph{Phys. Lett.}
  {\bfseries 104B} (1981) 199--202}.

\bibitem{Zhitnitsky:1980tq}
A.~R. Zhitnitsky, \emph{{On Possible Suppression of the Axion Hadron
  Interactions. (In Russian)}}, {\emph{Sov. J. Nucl. Phys.} {\bfseries 31}
  (1980) 260}.

\bibitem{Knirck:2019eug}
S.~Knirck, J.~Sch\"utte-Engel, A.~Millar, J.~Redondo, O.~Reimann, A.~Ringwald
  et~al., \emph{{A First Look on 3D Effects in Open Axion Haloscopes}},
  \href{https://doi.org/10.1088/1475-7516/2019/08/026}{\emph{JCAP} {\bfseries
  08} (2019) 026}, [\href{https://arxiv.org/abs/1906.02677}{{\ttfamily
  1906.02677}}].

\bibitem{Schutte-Engel:2018mfn}
{\scshape MADMAX} collaboration, J.~Schütte-Engel, \emph{{Simulation studies
  for the MADMAX axion direct detection experiment}},  in \emph{{14th Patras
  Workshop on Axions, WIMPs and WISPs (AXION-WIMP 2018) (PATRAS 2018) Hamburg,
  Germany, June 18-22, 2018}}, 2018,
  \href{https://arxiv.org/abs/1811.00493}{{\ttfamily 1811.00493}}.

\bibitem{Knirck:2020}
S.~Knirck, ``{How To Search for Axion Dark Matter with MADMAX (MAgnetized Disk
  and Mirror Axion eXperiment)}.'' dissertation, Technical University of
  Munich, Munich, 2020, \url{
  http://nbn-resolving.de/urn/resolver.pl?urn:nbn:de:bvb:91-diss-20200703-1542538-1-6}.

\bibitem{SchtteEngel:450146}
J.~Schütte-Engel, ``{A}xion direct detection in particle and condensed matter
  physics.'' dissertation, Universität Hamburg, Hamburg, 2020,
  \href{https://doi.org/10.3204/PUBDB-2020-04202}{10.3204/PUBDB-2020-04202}.

\bibitem{Redondo:2010js}
J.~Redondo, \emph{{Photon-Axion conversions in transversely inhomogeneous
  magnetic fields}},  in \emph{{Proceedings, 5th Patras Workshop on Axions,
  WIMPs and WISPs (AXION-WIMP 2009): Durham, UK, July 13-17, 2009}},
  pp.~185--188, 2010, \href{https://arxiv.org/abs/1003.0410}{{\ttfamily
  1003.0410}},
  \href{https://doi.org/10.3204/DESY-PROC-2009-05/redondo_javier}{DOI}.

\bibitem{Ouellet:2018nfr}
J.~Ouellet and Z.~Bogorad, \emph{{Solutions to Axion Electrodynamics in Various
  Geometries}}, \href{https://doi.org/10.1103/PhysRevD.99.055010}{\emph{Phys.
  Rev. D} {\bfseries 99} (2019) 055010},
  [\href{https://arxiv.org/abs/1809.10709}{{\ttfamily 1809.10709}}].

\bibitem{yeh2008essence}
C.~Yeh and F.~I. Shimabukuro, \emph{The essence of dielectric waveguides}.
\newblock Springer, 2008.

\bibitem{snyder1983optical}
A.~Snyder and J.~Love, \emph{Optical Waveguide Theory}.
\newblock Springer US, 1983.

\bibitem{goldsmith1998quasioptical}
P.~Goldsmith, \emph{Quasioptical Systems: Gaussian Beam Quasioptical
  Propogation and Applications}.
\newblock Wiley-IEEE Press, 1998,
  \href{https://doi.org/10.1109/9780470546291}{10.1109/9780470546291}.

\bibitem{Beurthey:2020yuq}
S.~Beurthey et~al., \emph{{MADMAX Status Report}},
  \href{https://arxiv.org/abs/2003.10894}{{\ttfamily 2003.10894}}.

\bibitem{Bergermann:2019}
D.~Bergermann, ``{Influence of outer fields on simulated MADMAX axion
  signals}.'' Bachelor thesis, Aachen, 2019.

\bibitem{Egge:2020hyo}
J.~Egge, S.~Knirck, B.~Majorovits, C.~Moore and O.~Reimann, \emph{{A first
  proof of principle booster setup for the MADMAX dielectric haloscope}},
  \href{https://doi.org/10.1140/epjc/s10052-020-7985-8}{\emph{Eur. Phys. J. C}
  {\bfseries 80} (2020) 392},
  [\href{https://arxiv.org/abs/2001.04363}{{\ttfamily 2001.04363}}].

\bibitem{Millar:2017eoc}
A.~J. Millar, J.~Redondo and F.~D. Steffen, \emph{{Dielectric haloscopes:
  sensitivity to the axion dark matter velocity}},
  \href{https://doi.org/10.1088/1475-7516/2017/10/006,
  10.1088/1475-7516/2018/05/E02}{\emph{JCAP} {\bfseries 1710} (2017) 006},
  [\href{https://arxiv.org/abs/1707.04266}{{\ttfamily 1707.04266}}].

\bibitem{Knirck:2018knd}
S.~Knirck, A.~J. Millar, C.~A.~J. O'Hare, J.~Redondo and F.~D. Steffen,
  \emph{{Directional axion detection}},
  \href{https://doi.org/10.1088/1475-7516/2018/11/051}{\emph{JCAP} {\bfseries
  1811} (2018) 051}, [\href{https://arxiv.org/abs/1806.05927}{{\ttfamily
  1806.05927}}].

\bibitem{Jaeckel:2013sqa}
J.~Jaeckel and J.~Redondo, \emph{{An antenna for directional detection of WISPy
  dark matter}},
  \href{https://doi.org/10.1088/1475-7516/2013/11/016}{\emph{JCAP} {\bfseries
  1311} (2013) 016}, [\href{https://arxiv.org/abs/1307.7181}{{\ttfamily
  1307.7181}}].

\bibitem{Jaeckel:2015kea}
J.~Jaeckel and S.~Knirck, \emph{{Directional Resolution of Dish Antenna
  Experiments to Search for WISPy Dark Matter}},
  \href{https://doi.org/10.1088/1475-7516/2016/01/005}{\emph{JCAP} {\bfseries
  1601} (2016) 005}, [\href{https://arxiv.org/abs/1509.00371}{{\ttfamily
  1509.00371}}].

\bibitem{Jaeckel:2017sjb}
J.~Jaeckel and S.~Knirck, \emph{{Dish Antenna Searches for WISPy Dark Matter:
  Directional Resolution Small Mass Limitations}},  in \emph{{Proceedings, 12th
  Patras Workshop on Axions, WIMPs and WISPs (PATRAS 2016): Jeju Island, South
  Korea, June 20-24, 2016}}, pp.~78--81, 2017,
  \href{https://arxiv.org/abs/1702.04381}{{\ttfamily 1702.04381}},
  \href{https://doi.org/10.3204/DESY-PROC-2009-03/Knirck_Stefan}{DOI}.

\bibitem{Halpern:86}
M.~Halpern, H.~P. Gush, E.~Wishnow and V.~D. Cosmo, \emph{Far infrared
  transmission of dielectrics at cryogenic and room temperatures: glass,
  fluorogold, eccosorb, stycast, and various plastics},
  \href{https://doi.org/10.1364/AO.25.000565}{\emph{Appl. Opt.} {\bfseries 25}
  (Feb, 1986) 565--570}.

\bibitem{LALOsize:website}
\url{http://www.crystec.de/daten/laalo3.pdf} [accessed 2020-11-23].

\bibitem{Barto:2017}
A.~Barto, S.~Winters, J.~Burge, D.~Davies, H.~Doty, R.~Richer et~al.,
  \emph{{Design and component test results of the LSST Camera L1-L2 lens
  assembly}},  in \emph{Astronomical Optics: Design, Manufacture, and Test of
  Space and Ground Systems} (T.~B. Hull, D.~W. Kim and P.~Hallibert, eds.),
  vol.~10401, pp.~125 -- 131, International Society for Optics and Photonics,
  SPIE, 2017, \href{https://doi.org/10.1117/12.2274808}{DOI}.

\bibitem{Planat_2020}
L.~Planat, A.~Ranadive, R.~Dassonneville, J.~Puertas~Martínez, S.~Léger,
  C.~Naud et~al., \emph{Photonic-crystal josephson traveling-wave parametric
  amplifier}, \href{https://doi.org/10.1103/physrevx.10.021021}{\emph{Physical
  Review X} {\bfseries 10} (Apr, 2020) }.

\bibitem{Rybka:2014cya}
G.~Rybka, A.~Wagner, A.~Brill, K.~Ramos, R.~Percival and K.~Patel,
  \emph{{Search for dark matter axions with the Orpheus experiment}},
  \href{https://doi.org/10.1103/PhysRevD.91.011701}{\emph{Phys. Rev.}
  {\bfseries D91} (2015) 011701},
  [\href{https://arxiv.org/abs/1403.3121}{{\ttfamily 1403.3121}}].

\bibitem{cervantes2019orpheus}
R.~Cervantes et~al., \emph{{Orpheus: Extending the ADMX QCD Dark-Matter Axion
  Search to Higher Masses}}, {\emph{APS April Meeting} (2019) Z09--005}.

\bibitem{DeMiguel-Hernandez:2020mon}
J.~De~Miguel-Hernández, \emph{{A dark matter telescope probing the 6 to 60 GHz
  band}},  \href{https://arxiv.org/abs/2003.06874}{{\ttfamily 2003.06874}}.

\end{thebibliography}\endgroup

\end{document}